\documentclass[a4paper,11pt,bookmarks=true]{article}
\pdfoutput=1
\usepackage{jheppub}
\usepackage[colorlinks=true,linkcolor=blue,citecolor=blue, urlcolor=blue]{hyperref}   
\usepackage{xcolor}
\usepackage{graphicx}
\usepackage{dcolumn}
\usepackage{bm}
\usepackage{amsmath}

\newcommand{\p}{\bm{p}}
\newcommand{\x}{\bm{x}}
\newcommand{\n}{\bm{n}}
\def\p{\bm{p}}
\def\x{\bm{x}}
\def\k{\bm{k}}
\def\pp{\bm{p'}}
\def\kp{\bm{k'}}
\def\ptilde{\tilde{\p}}
\usepackage[capitalise]{cleveref}

\begin{document}

\title{Minijet quenching in non-equilibrium quark-gluon plasma }

\author[a]{Fabian Zhou}%
\emailAdd{f.zhou@thphys.uni-heidelberg.de}
\affiliation[a]{{}Institute for Theoretical Physics, University of Heidelberg, 69120 Heidelberg, Germany}

\author[b]{Jasmine Brewer}
\emailAdd{jasmine.brewer@physics.ox.ac.uk}
\affiliation[b]{Rudolf Peierls Centre for Theoretical Physics, University of Oxford, Oxford OX1 3PU, UK}

\author[a]{Aleksas Mazeliauskas}
\emailAdd{a.mazeliauskas@thphys.uni-heidelberg.de}

\date{\today}

\abstract{
    We study the energy deposition and thermalisation of high-momentum on-shell partons (minijets) travelling through a non-equilibrium Quark-Gluon Plasma using QCD kinetic theory. For thermal backgrounds, we show that the parton energy first flows to the soft sector by collinear cascade and then isotropises via elastic scatterings.
    In contrast, the momentum deposition from a minijet reaches the equilibrium distribution directly. 
 For expanding non-equilibrium QGP, we study the time for a minijet perturbation to lose memory of its initial conditions, namely, the hydrodynamisation time. We show that the minijet evolution scales well with the relaxation time $\tau_R\propto \eta/s/T(\tau)$, where $T(\tau)$ is the effective temperature and $\eta/s$ is the viscosity over entropy ratio. 
}

    \maketitle

\tableofcontents

\section{Introduction}

High-energy collisions of heavy ions at the Relativistic Heavy Ion Collider (RHIC) at BNL and the Large Hadron Collider (LHC) at CERN have produced ample evidence for the creation of a hot and dense medium consisting of deconfined quarks and gluons---the Quark Gluon Plasma (QGP). One of the key signatures of the formation of the QGP is the suppression of high momentum particles produced in the collision compared to the expectation from proton-proton collisions~\cite{STAR:2005gfr,PHENIX:2004vcz,Aad:2010bu,Chatrchyan:2011sx,Adam:2015ewa} (see ~\cite{Connors:2017ptx,Apolinario:2022vzg,Cunqueiro:2021wls} for reviews). This phenomenon, called jet quenching, is interpreted as energy loss of high-momentum quarks and gluons when they travel through the QCD medium produced in heavy-ion collisions. The quenching of jets has the potential to provide a powerful probe of the microscopic structure of the QGP and its macroscopic transport properties.

Initially, hard scatterings in the collision create highly energetic partons. Because of the large virtuality of these partons, vacuum-like splittings happen much faster than typical timescales for the medium interactions. Therefore, at this stage the evolution of a shower is vacuum-like. The parton sheds virtuality by splitting into softer and less virtual partons. Once the partons become on-shell, there can only be splittings triggered by interactions with the medium. We call such on-shell partons \emph{minijets} to distinguish them from high virtuality partons undergoing a vacuum shower evolution\footnote{A closely related concept of minijets~\cite{Kajantie:1987pd} refers instead to semi-hard momentum partons produced in the initial hard scattering.}.

Extensive theoretical and phenomenological works have focused on describing parton energy loss in the QGP.
In a perturbative QCD picture, a parton travelling through the QCD medium interacts with it by scattering with the background gluon fields. The modifications to a high energy parton and its splitting probabilities can be computed in the BDMPS-Z framework~\cite{Baier:1996sk,Baier:1996kr,Zakharov:1996fv,Zakharov:1997uu}.
Semi-analytical approaches based on this framework describe jet quenching as the multiple scattering of a QCD parton or single vacuum-like splitting~\cite{Casalderrey-Solana:2011ule}, typically in a QCD medium that is a static ``brick'' (see~\cite{Blaizot:2015lma} for a review), but more recently also undergoing longitudinal expansion~\cite{Andres:2023jao}, interacting with a highly-anisotropic medium~\cite{Hauksson:2021okc,Hauksson:2023tze}, or even including the flow of the plasma~\cite{Barata:2023qds,Kuzmin:2023hko}. Phenomenological approaches, for example~\cite{Zapp:2013vla}, interface the vacuum-like evolution of a jet (via a parton shower) with perturbative scatterings with the QCD medium. In~\cite{Caucal:2018dla, Caucal:2019uvr} 
this is done through the explicit factorisation of medium effects from the perturbative, vacuum-like splittings.
In \cite{Mehtar-Tani:2021fud, Mehtar-Tani:2024jtd} the jet nuclear modification factor is computed semi-analytically by resumming the energy loss of resolved vacuum-like splittings.
Many other models~\cite{Casalderrey-Solana:2014bpa,He:2015pra,JETSCAPE:2021ehl,Schenke:2009gb} have more extensive phenomenological success, but in turn require an interface between a parton shower and the medium described by hydrodynamics, which no longer carries information about individual QCD partons in the medium.
In addition to extensive phenomenological extractions~\cite{JET:2013cls,Andres:2016iys,Andres:2019eus,Huss:2020whe,JETSCAPE:2021ehl}, there have also been several theoretical calculations of jet quenching parameter $\hat{q}$ in various settings.
Notably, this includes equilibrium calculations using lattice EQCD simulations~\cite{Moore:2021jwe,Moore:2021jwe}, QCD kinetic theory at NLO~\cite{Ghiglieri:2015ala} and hadron transport models~\cite{Dorau:2019ozd}. At early non-equilibrium phases of heavy-ion collision $\hat{q}$ is computed from strong glasma fields~\cite{Ipp:2020mjc,Ipp:2020nfu,Carrington:2021dvw,Carrington:2022bnv,Avramescu:2023qvv} and non-equilibrium QCD kinetic theory~\cite{Boguslavski:2023alu,Boguslavski:2023waw}.

As a high-energy parton interacts with, and loses energy to, the QCD medium, that energy and momentum are transferred to the medium itself. This phenomenon of medium response seems to be crucial for phenomenological descriptions of measurements on jets~\cite{Casalderrey-Solana:2016jvj,Milhano:2017nzm, Tachibana:2017syd,KunnawalkamElayavalli:2017hxo} and particularly impacts relatively soft particles at large angles from the jet axis. Several works have focused on developing jet observables that aim to reduce the effect of medium response to gain better perturbative control~\cite{Mehtar-Tani:2016aco,Casalderrey-Solana:2019ubu, Mehtar-Tani:2019rrk,Mulligan:2020tim,Apolinario:2020uvt,Caucal:2021cfb,Cunqueiro:2023vxl}, or to highlight its effects for improved sensitivity to physics in this regime~\cite{Casalderrey-Solana:2016jvj,Milhano:2017nzm,Pablos:2019ngg,Brewer:2021hmh,Yang:2021qtl}.
However, this is precisely the regime in which it is challenging to treat the backreaction of the medium due to energy and momentum lost by high-energy partons
using semi-analytical descriptions.
Models like the dynamical core-corona initialisation~\cite{Kanakubo:2019ogh,Kanakubo:2021qcw}, the minijet+hydro framework~\cite{Pablos:2022piv}, the CoLBT-hydro model~\cite{Chen:2017zte}, the hybrid model~\cite{Casalderrey-Solana:2020rsj}, and \textsc{Jetscape}~\cite{JETSCAPE:2022tgg} include hard partons as source terms in the hydrodynamic equations of motion which enables them to study the feedback of jets on the evolution of the background medium.
However, in these studies, the background QGP and jet perturbations are treated in different frameworks, and therefore they must introduce some transition from the perturbative showering regime to the hydrodynamic description. In the EKRT and McDIPPER models for the initial conditions of hydrodynamic evolution, the energy density profile is obtained from the perturbative production of few GeV partons~\cite{Niemi:2015qia,Eskola:1999fc,Garcia-Montero:2023gex}. In this approach, energy of minijets is directly input to hydrodynamics without a dynamic description of thermalisation. The goal of this work is to study the energy loss and thermalisation of high-momentum partons in the out-of-equilibrium QGP in a framework that naturally includes both high-momentum partons and QCD partons in the medium. 

In the high-energy limit, the bulk properties and thermalisation of QCD medium can be described by an Effective Kinetic Theory (EKT) of QCD interactions~\cite{Arnold:2002zm}. At leading order in the coupling constant, the massless gluon and quark degrees of freedom evolve according to Boltzmann equations with elastic scattering and collinear radiation processes. The interplay of these collision processes with rapid longitudinal expansion in heavy-ion collisions leads to the celebrated ``bottom-up" thermalisation picture~\cite{Baier:2000sb, Berges:2020fwq,Schlichting:2019abc}. Numerical implementations of EKT have yielded a detailed description of QGP equilibration and hydrodynamisation in heavy-ion collisions~\cite{Kurkela:2014tea,Kurkela:2015qoa,Kurkela:2018oqw,Kurkela:2018xxd,Du:2020dvp,Du:2020zqg}.
Physically, the energy loss of high-momentum partons is driven by the same scattering processes as the equilibration of the QGP. 
The separation between the medium and a jet is merely a separation in the energy scale. A jet transfers energy to softer and softer partons that eventually are no longer parametrically separated from the medium. Therefore, the final stages of jet thermalisation are best described in the common framework of EKT. In this work, we use QCD kinetic theory to describe the evolution of high-energy and on-shell partons (minijets) and their thermalisation with the bulk QGP.

Recently, there were several works studying minijet thermalisation using kinetic theory~\cite{Iancu:2015uja,Mehtar-Tani:2018zba,Schlichting:2020lef,Mehtar-Tani:2022zwf,Sirimanna:2022zje}.
It has been shown that the energy of the partons is transported down to the medium temperature scale via a turbulent cascade. The energy that escapes the jet cone is mostly transported by the soft fragments of the in-medium cascade. 
However, these works considered minijets propagating in a thermal background. In heavy-ion collisions, the rapid longitudinal expansion introduces significant momentum anisotropies in the momentum distribution of QGP particles, which is especially relevant at the earliest stages of the collision. In this work, we extend the previous studies of minijet thermalisation to anisotropic and expanding backgrounds.

The paper is organised as follows. In \cref{setup}, we introduce the effective kinetic description of QCD and motivate why linearised equations are suitable to study minijets. 
We also specify the initial conditions for the background and minijets. Next, in \cref{non_exp_plasma}, we present a systematic study of jet thermalisation with an increasing complexity of the background. We consider the case of a non-expanding medium. For thermal backgrounds, this reproduces previous results, where we clarify the angular dependence of minijet thermalisation and introduce the concept of angle-dependent temperature. We also study minijet thermalisation in an anisotropic background without expansion. 
In \cref{exp_plasma}, we consider the longitudinally expanding case relevant to heavy-ion phenomenology. We study the time it takes for minijets to become part of the background, i.e., for them to hydrodynamise.  Our conclusions are given in \cref{sec:concl}.
In \cref{lin_kernels}, we provide explicit formulas for the linearised collision kernel.

\section{Setup}\label{setup}

\subsection{QCD kinetic theory}\label{sec:kinetic_theory}
The framework we are working with is the QCD kinetic theory (AMY)~\cite{Arnold:2002zm} which is a leading order description in the coupling $\lambda = N_c g^2$. As implementation details have been discussed in multiple previous publications~\cite{Kurkela:2014tea,Kurkela:2015qoa,Kurkela:2018oqw,Kurkela:2018xxd,Du:2020dvp,Du:2020zqg}, here we will provide only a short summary.

In the high-temperature (conformal) limit of QCD, the dominant degrees of freedom for describing the energy-momentum tensor are shown to be massless quark and gluon quasi-particles. Generally, the phase-space distribution function $f_s(t,\x,\p)$ of some particle species $s$ obeys a Boltzmann equation
\begin{equation}
    \left(\partial_{t} + \hat{\bm{p}}\cdot \nabla_{\x}\right)f_s(\tau,\x,\p) = -C^s[f],
\end{equation}
where $\hat{\bm{p}}$ is a unit vector. The collision kernel $C^s[f]$ contains all QCD inelastic 1$\leftrightarrow$2 and elastic 2$\leftrightarrow$2 processes~\cite{Arnold:2002zm}, 
\begin{equation}
    C^s[f] = C^s_{2\leftrightarrow2}[f] + C^s_{1\leftrightarrow2}[f].
\end{equation}
The elastic kernel $C^s_{2\leftrightarrow2}$ is given by
\begin{equation}
\begin{aligned}\label{eq:C2to2}
C^s_{2\leftrightarrow 2}[f](\ptilde) &= \frac{1}{2}
\frac{1}{\nu_s} \frac{1}{4}
 \sum_{abcd} (2\pi)^3 \int_{\p\k\p'\k'}
 |\mathcal{M}^{ab}_{cd} 
 |^2(2\pi)^4
 \delta^{(4)}(P+K-P'-K')\\
 &\times\{ (f^a_{\p} f^b_{\k} (1\pm f^c_{\p'})(1\pm f^d_{\k'}))-(f^c_{\p'} 
f^d_{\k'} (1\pm f^a_{\p})(1 \pm f^b_{\k})) \}\\
&\times 
\left[\delta(\ptilde-\p)\delta_{as}+\delta(\ptilde-\k)\delta_{bs} 
-\delta(\ptilde-\p')\delta_{cs}-\delta(\ptilde-\k')\delta_{ds}\right],
\end{aligned}
\end{equation}
where we integrate over the in-coming and out-going particle momenta with the Lorentz invariant measure $\int_{\p} = \int\frac{d^3\p}{(2\pi)^{3}2p}$. Here $|\mathcal{M}^{ab}_{cd}|^2$ represents the squared scattering amplitude for $ab\leftrightarrow cd$ process summed over spin/polarisation and colour degrees of freedom $\nu_s$ for each in-coming and out-going particle ($\nu_s$ is 16 for gluons and 6 for each quark/anti-quark flavour). The Dirac-delta function in the first line ensures energy and momentum conservation, while the distribution functions in the second line represent the usual loss and gain terms. The sum over external particles $\sum_{abcd}$ with the four terms in the last line sums over all possibilities for the particle $s$ with momentum $\ptilde$ to participate in a  $2\leftrightarrow2$ scattering.
For large momentum transfer, the matrix elements $|\mathcal{M}^{ab}_{cd}|^2$ coincide with the tree level $2\leftrightarrow2$ processes of QCD~\cite{Arnold:2002zm}. In the case of purely gluonic scattering, we have
\begin{equation}
    |\mathcal{M}^{gg}_{gg}|^2 = 2\lambda^2\nu_g\left( 9+ \frac{(s-t)^2}{u^2} + \frac{(u-s^2)}{t^2} + \frac{(t-u)^2}{s^2} \right).
\end{equation}
The $t$ channel has an infrared divergence for small momentum transfer $q = |\p-\p'|$, with incoming momentum $\p$ and outgoing $\p'$ (similarly for the $u$ channel), which is regulated by a medium-induced effective mass $m_{s}^2$, where
\begin{align}
\label{eq:effective_mass_g}
    m_g^2 &= 4 g^2\int_{\p}[N_c f_g(p) + \frac{N_f}{2} ( f_q(p) + f_{\Bar{q}}(p))],\\
\label{eq:effective_mass_q}
    m_q^2 &= 4 g^2C_F\int_{\p}[2 f_g(p) + \frac{N_f}{2} ( f_q(p) + f_{\Bar{q}}(p))].
\end{align}
A commonly used prescription is an isotropic screening which replaces $q^2 \rightarrow q^2 + \xi_{s}^2 m_{s}^2$, where
the coefficients $\xi_g = e^{5/6}/2$ and $\xi_q = e/2$ are chosen to reproduce the full HTL results for the gluon drag and momentum diffusion properties of soft gluon scattering and gluon to quark conversion in thermal equilibrium~\cite{AbraaoYork:2014hbk,Ghiglieri:2015ala,Kurkela:2018oqw,Berges:2020fwq}.

The inelastic kernel $C_{1\leftrightarrow2}$ describes the medium-induced collinear radiation of gluon bremsstrahlung as well as splittings into quark-antiquark pairs
\begin{equation}
\begin{aligned}
C^s_{1\leftrightarrow 2}[f](\ptilde) &=\frac{1}{2}
\frac{1}{\nu_s} \frac{(2\pi)^3}{4\pi\tilde{p}}
 \sum_{abc}\int_0^{\infty}dpdp'dk' 
 4\pi\gamma^{a}_{bc} 
 \delta(p-p'-k')\\
&\times \{ (f^a_{p\hat{\bm{n}}} (1\pm f^b_{p'\hat{\bm{n}}})(1\pm f^c_{k'\hat{\bm{n}}}))-(f^b_{p'\hat{\bm{n}}} 
f^c_{k'\hat{\bm{n}}} (1\pm f^a_{p\hat{\bm{n}}})\} \\
&\times\left[\delta(\Tilde{p}-p)\delta_{as}-\delta(\tilde{p}-p')\delta_{bs} 
-\delta(\tilde{p}-k')\delta_{cs}\right].
    \end{aligned}
\end{equation}
The splitting rates $\gamma^{a}_{bc}$ satisfy an integral equation, which has to be solved self-consistently to incorporate multiple interactions with the medium (see \cref{lin_kernels}). The unit vector $\hat{\bm{n}} = \tilde{\p}/|\tilde{\p}|$ points in the splitting direction. Because the formation time for emitted radiation grows with energy~\cite{Arnold:2002zm}, separate medium scatterings interfere, leading to the so-called Landau-Pomeranchuk-Migdal suppression~\cite{Landau:1953um,Landau:1953gr,Migdal:1956tc,Arnold:2002ja}.

For early times in a heavy ion collision, the dynamics is dominated by the rapid expansion of the system in the longitudinal direction. This allows us to approximate the system as being homogeneous in the transverse plane and boost-invariant in the longitudinal direction. The evolution of the initially far-from-equilibrium system is then described by a Boltzmann equation
\begin{equation}\label{eq:hom_boltzmann}
    \left(\partial_{\tau} - \frac{p_z}{\tau}\partial_{p_z}\right)f_s(\tau,\p) = -C^s[f],
\end{equation}
where $\tau=\sqrt{t^2-z^2}$. The second term on the left-hand side originates from gradients in $z$-direction, which are written as gradients in momentum space using boost-invariance~\cite{Baym:1984np}.

In this work, we focus on the thermalisation of minijets inside the medium, which we model as a single energetic parton going through a medium. 
We decompose $f$ into background and minijet (mj) perturbation
\begin{equation}
    f_s(\tau,\p) = \Bar{f}_s(\tau,\p) + \delta f_{s,\text{mj}}(\tau,\p).
\end{equation}
Note that we do not have a spatial coordinate dependence for $\delta f_\text{jet}$, which can be considered a local perturbation in momentum space, but homogeneous in coordinate space. An actual minijet particle is local both in momentum and coordinate space, therefore, our study is limited to understanding the equilibration in momentum space and cannot tell how the jet wake develops in coordinate space. Implementing spatial localisation is important but computationally demanding and we leave it for future work.

The QGP fireball mainly consists of soft particles and the occupancies of high momentum partons produced are much smaller than that of the background partons, i.e., $\delta f_{\textrm{mj}}\ll \Bar{f}$. In addition, we can assume the high-momentum partons to carry a small fraction of the total energy.
This enables us to linearize the equations of motion, leading to a set of coupled equations  for $\bar{f}_s(\tau,\p)$ and $\delta f_s(\tau,\p)$
\begin{subequations}\label{eq:linearised}
\begin{align}
    \left(\partial_{\tau} - \frac{p_z}{\tau}\partial_{p_z}\right)\Bar{f}_s(\tau,\p) &= - C^s[\Bar{f}],\label{eq:linearised1}\\
    \left(\partial_{\tau} - \frac{p_z}{\tau}\partial_{p_z}\right)\delta f_s(\tau,\p) &= -\delta C^s[\Bar{f},\delta f].\label{eq:linearised2}
\end{align}
\end{subequations}
We note that we keep the full non-linear evolution of the background distribution in \cref{eq:linearised1}.
The linearised collision kernel $\delta C^s[\Bar{f},\delta f]$ describes how the background affects the evolution of $\delta f_s(\tau,\p)$. Except when studying chemical equilibration, we will drop the species subscript for simplicity. Linearising $C_{2\leftrightarrow 2}$, \cref{eq:C2to2}, we receive one contribution from the loss and gain terms and one from the perturbation of the effective mass $m_{g,q}$ (\cref{eq:effective_mass_g,eq:effective_mass_q}) that regulate the matrix elements. The inelastic kernel $\delta C_{1\leftrightarrow 2}$ also receive contributions from linearising loss-gain terms and the splitting rate $\gamma^{a}_{bc}$. We summarise the explicit expressions of the linearised kernels in \cref{lin_kernels}.

 Because we work with continuous distributions in phase space, linearisation enforces the physical condition that a parton can not scatter from itself. Although there is no back-reaction of the perturbation onto the background distribution $\bar f$, the perturbation $\delta f$ eventually equilibrates and $\bar f +\delta f$ can be considered as a new background. To derive \cref{eq:linearised2} we formally require $\delta f\ll 1$ and $\delta f \ll \bar{f}$, but once the equations are linearised the magnitude of $\delta f$ and even the sign can be arbitrary. 

The thermalisation of background gluon and quark distributions by solving \cref{eq:linearised1} has been studied extensively in previous works~\cite{Kurkela:2014tea,Kurkela:2015qoa,Kurkela:2018oqw,Kurkela:2018xxd,Du:2020dvp,Du:2020zqg}.
The linearised kinetic equations of Yang-Mills (YM) theory have been solved to derive linear non-equilibrium response functions for medium perturbations at the characteristic medium energy scale~\cite{Keegan:2016cpi,Kurkela:2018vqr,Kurkela:2018wud}. The equilibration of high momentum perturbations in gluon-only kinetic theory has been studied previously for the isotropic static background in~\cite{Kurkela:2014tea}. More recently, the linearised QCD equations \cref{eq:linearised} have been solved for high-momentum perturbations around a static thermal background in Ref.~\cite{Schlichting:2020lef,Mehtar-Tani:2022zwf}. We extend these studies, in particular, by considering high-momentum perturbations in QCD kinetic theory with non-thermal and expanding backgrounds.

\subsection{Initial conditions}

\subsubsection{Background}

In thermal equilibrium, the quark and gluon distributions are uniquely defined by a single energy scale given by temperature $T$, and the function
\begin{equation}\label{therm_distr}
    f_{\mathrm{th}}(p) = \frac{1}{e^{p/T}\pm 1},
\end{equation}
where $(-)$ corresponds to the Bose-Einstein distribution for gluons and $(+)$ corresponds to the Fermi-Dirac distribution for quarks.
The uniqueness of the temperature in a thermal distribution is given by the fact that it can be extracted from any moment of the distribution function
\begin{equation}
I_n= \int\! \frac{d^3 \p}{(2\pi)^3}  p^n f_{\mathrm{th}}(p) = \mathcal{N}_{n,\pm} T^{n+3}_{n},
\label{eq:Teff}
\end{equation}
where the normalisation constant is
\begin{equation}
\mathcal{N}_{n, \pm} = [1-2^{-n-2}]^{\frac{1\pm 1}{2}} \frac{\Gamma(n+3)\zeta(n+3)}{2\pi^2 }.
\end{equation}

In the special case of a thermal background in \cref{eq:Teff}, $T_n = T$ for all $n$, which is just the temperature of the system. Nevertheless, this equation can be generalised for arbitrary backgrounds $\Bar{f}$. In that case it is not guaranteed that the different $T_n$ agree. Conventionally, one considers $n=1$ to obtain the effective temperature $\overline{T} = T_1$ that we get from the Landau matching of the energy density, i.e.
\begin{equation}\label{eq:e_landau_matching}
    \Bar{e}(\tau) =\nu_\text{eff} \frac{\pi^2}{30}\overline{T}(\tau)^4\, .
\end{equation}
Here $\nu_{\textrm{eff}} = \nu_g+ \frac{7}{8}\nu_q$ are the effective degrees of freedom with $\nu_g = 16$ for gluons and $\nu_q = 36$ for 3 flavours of light quarks and anti-quarks. Note that we consider the QGP at zero baryon chemical potential and quark and anti-quark distributions are assumed to be equal.

In the high-energy limit, the energy deposition in the mid-rapidity region is dominated by the scattering of small Bjorken-$x$ gluons~\cite{Berges:2020fwq}. This leads to gluon saturation phenomena with high gluon occupancies $f_g\sim\mathcal{O}(\lambda^{-1})\gg1$ up to saturation scale $Q_s$. 
The initial gluon distribution is highly anisotropic, with typical longitudinal momentum much smaller than the transverse momentum $\left<p_z^2\right> \ll \left<p_T^2\right>\sim Q_s^2$, where $p_T^2 = p_x^2 + p_y^2$. We will use the previously used CGC-parametrisation of an overoccupied gluonic plasma~\cite{Kurkela:2015qoa}
\begin{equation}\label{init_prl_distr}
    f_\text{sat} (p_T, p_z) = \frac{2A}{\lambda}\frac{Q_0e^{-\frac{2}{3}\frac{1}{Q_0^2}(p_T^2 + \xi^2 p_z^2)}}{\sqrt{p_T^2 + \xi^2 p_z^2}},
\end{equation}
where the parameter $\xi=10$ accounts for the anisotropy of the initial distribution and $Q_0=1.8Q_s$ sets the momentum scale of the background. The constant $A=5.24$ is fixed such that the comoving energy density $\tau e$ agrees with classical lattice simulations. The initial density of quarks is set to zero. During the evolution, the quarks will be produced dynamically and the system will undergo both chemical and kinetic equilibration~\cite{Kurkela:2018oqw,Kurkela:2018xxd}.

\subsubsection{Minijet perturbation}

\begin{figure}
\centering
    \includegraphics[width=0.6\columnwidth]{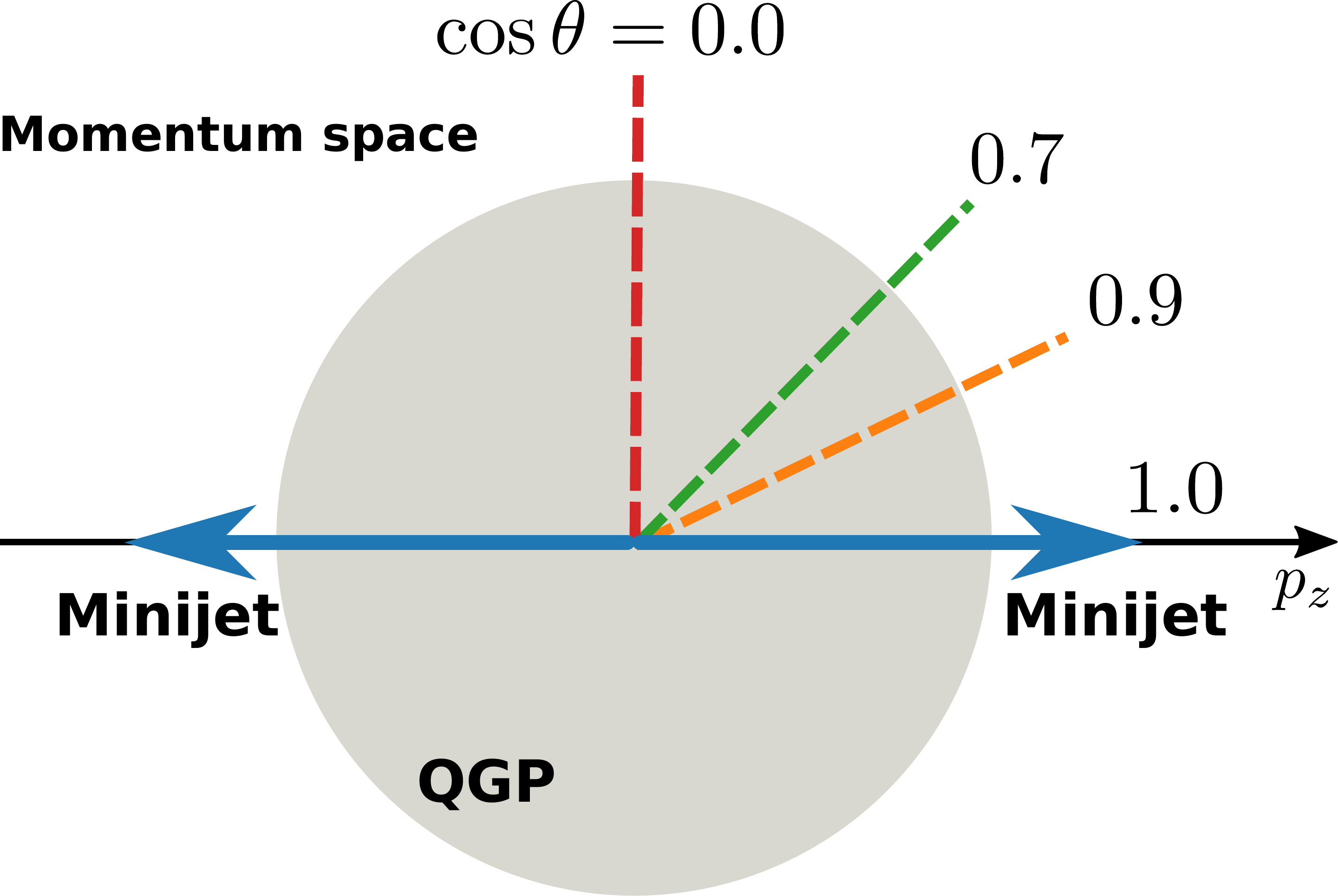}
    \caption{Illustration of a parity-even minijet configuration in momentum space for a thermal QGP. The background is isotropic and we choose the perturbation to point in $z$-direction.\label{fig:dijetcartoon}}
\end{figure}

A single minijet $\delta f(\p) = (2\pi)^3 \delta^{(3)}(\p-E \hat{\n})$ is uniquely characterised by its energy $E$ and direction of propagation $\hat{\n}$\footnote{As stated in \cref{sec:kinetic_theory}, we consider perturbations homogeneous in space.}. In practice, we approximate the perturbation by a Gaussian of width $\sigma=0.1 E$\footnote{We checked that for a smaller width of $\sigma = 0.05E$ there is no difference in the isotropisation of $\delta f(\p)$.}. For a minijet in $z$-direction the distribution is written as
\begin{equation}\label{initial_gluon_distr}
\delta f(\p) = \left(\frac{2\pi}{\sigma^2}\right)^{3/2}\frac{E}{p} \exp\left(-\frac{ p_x^2+ p_y^2+ (p_z - E)^2 }{2\sigma^2}\right).
\end{equation}
From this expression, it is easy to see that the components of the perturbed energy-momentum tensor and number density (for a single degree of freedom) read
\begin{align}
\delta T^{tt} &=\int \frac{d^3\p}{(2\pi)^3} p \delta f =E ,\\
\delta T^{tz} &=\int \frac{d^3\p}{(2\pi)^3} p^z \delta f \approx E + \mathcal{O}\left(\frac{\sigma^2}{E^2}\right),\\
\delta n &=\int \frac{d^3\p}{(2\pi)^3} \delta f \approx 1+ \mathcal{O}\left(\frac{\sigma^2}{E^2}\right).
\end{align}
Note that the normalisation for linear perturbations is arbitrary, so these values can be rescaled by an arbitrarily small constant. If one considers more than one species of particles, e.g. gluons and (anti-)quarks, one has to perform the sum over all species $a$ weighted by the corresponding degrees of freedom $\nu_s$.

Kinetic evolution conserves energy and momentum. Therefore, an equilibrated minijet increases the thermal gluon or quark distribution by the corresponding energy and momentum. This corresponds to changing the temperature and boosting the thermal distribution.
Explicitly, the thermal distribution for a perturbation is given by
\begin{subequations}\label{eq:thjet}
\begin{align}
\delta f_{\textrm{th}}(\p) &=\left( \delta T \partial_T + \delta u^z \partial_{u^z} \right) f_{\textrm{th}}\left(p_{\mu}u^{\mu}/T\right)\bigg|_{u^z=0}\\
     &=\left( \frac{\delta T}{T} +  \delta \bm{v}\cdot\hat{\p}\right) \frac{p}{T} f_{\textrm{th}}(p)(1\pm f_{\textrm{th}}(p)),
\end{align}
\end{subequations}
where the 3-velocity is defined by $u^{\mu} = \gamma (1,\bm{v})$, $\gamma=1/\sqrt{1-v^2}$. From \cref{eq:thjet} one can straightforwardly compute the moments $\delta T^{tt}=\frac{\pi^2}{30}\nu_{\textrm{eff}}T^4 \frac{4\delta T}{T}$ and $\delta T^{tz}=\frac{\pi^2}{30}\nu_{\textrm{eff}}T^4 \frac{4\delta v}{3}$.

In equilibrium, the temperature perturbation is isotropic in momentum angle, while velocity perturbations have $\cos\theta$ modulation with respect to minijet direction. Respectively, these perturbations are even and odd under the parity transformation $\p \leftrightarrow -\p$. Parity-even and odd perturbations relate to different conserved quantities. Parity-even perturbations impact the net change in energy of the thermalised state but do not change the net momentum, while parity-odd perturbations do not change the energy but do add net momentum. To disentangle the evolution of these two contributions, we will separately study the evolution of parity even and odd initial conditions. We will call these contributions net-energy and net-momentum perturbations, respectively. 
Thanks to the linearity, the minijet and its subsequent evolution is a sum of these two configurations:
\begin{equation}\label{eq:jet_decomposition}
\delta f(\p)  = \underbrace{\frac{1}{2}(\delta f(\p) +\delta f(-\p))}_{\text{net-energy perturbation}}+\underbrace{\frac{1}{2}(\delta f(\p) -\delta f(-\p))}_{\text{net-momentum perturbation}}
\end{equation}
We emphasise that for momentum-integrated and parity-even quantities, like energy density or pressure, the parity-even (net-energy) perturbation is the only contribution to the minijet evolution. Similarly, momentum-integrated quantities that are parity-odd can be understood from the parity-odd (net-momentum) perturbations only. In \cref{fig:dijetcartoon} we show a cartoon of minijets in the net-energy configuration, in \emph{momentum} space. We emphasize that, throughout this work, we will study perturbations (minijets) in momentum space only. In the following, we will study how the minijet equilibration depends on the angle with respect to the minijet axis. In a thermal background, there is no preferred orientation of the minijet and we can choose the angle $\cos\theta =1$ to coincide with the minijet direction as shown in the figure. However, in the longitudinally expanding case, the $z$ direction is singled out as the beam axis. In this case, the direction $\cos\theta=1$  is along the $z$-axis and the direction of a minijet in the transverse plane is given by $\cos \theta =0$. We will assume that the background distribution is azimuthally-symmetric around the beam axis, so we can choose the minijet direction to be in the $x-z$ plane, i.e., $\phi=0$.  In the following, we will always consider a coupling of $\lambda = 10$ and minijet energy of $E=30\overline{T}$ (for constant-temperature backgrounds) or $E=30Q_s$ (for non-equilibrium backgrounds) unless otherwise stated.

\section{Minijet thermalisation in non-expanding plasma}\label{non_exp_plasma}

In this section, we study thermalisation of minijets on top of a background distribution without longitudinal expansion. Except for \cref{Anisotropic background}, the background is assumed to be thermal at temperature $\overline{T}$. We work in the usual Minkowski coordinates with time variable $t$.

\subsection{Equilibration of net-energy perturbations}

\subsubsection{Two stage thermalisation}

\begin{figure*}
    \centering
    \includegraphics[scale=1.0]{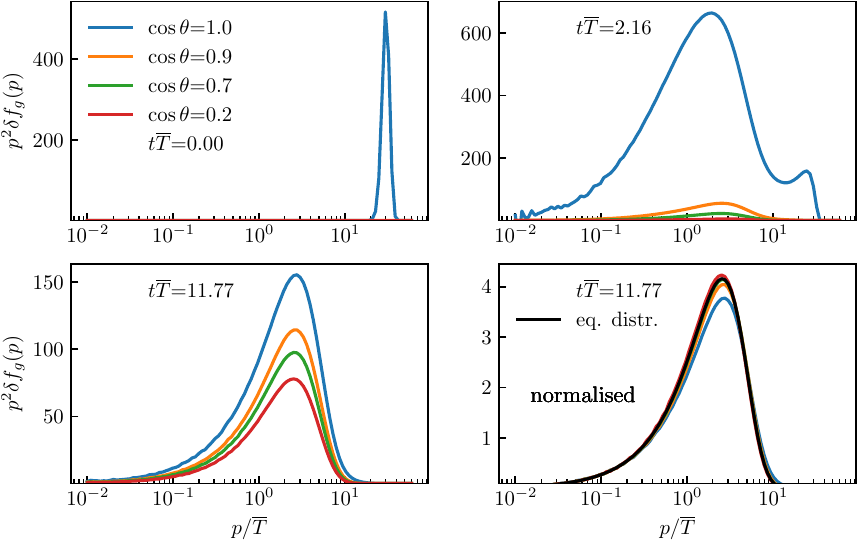}
    \caption{Distribution of $\delta f(p,\theta)$ for a net-energy perturbation as a function of time, with different colours corresponding to different angles $\cos\theta$. In the last panel, we scale the curves by the area underneath and compare them with the equilibrium distribution.}
    \label{fig:df_unscaled_d_E}
\end{figure*}

We first study purely gluonic plasma, i.e., YM kinetic theory, with $\lambda=10$. In this section, we focus on the perturbations that change the total energy in the thermalised state, namely, those that are even under ${\bf p} \rightarrow -{\bf p}$. In \cref{fig:df_unscaled_d_E} we show the evolution of such perturbations in different angular slices (cf. \cref{fig:dijetcartoon}) for different times. The initial perturbation is centred at $E=30 \overline{T}$, where $\overline{T}$ is the background temperature. The initial peak rapidly decays and the $p\sim \overline{T}$ region is populated by the time $t\overline{T}\approx 3$ but is still very anisotropic. For $t\overline{T}> 10$ the initial minijet peak is no longer visible and all angles are populated with the same shape of distribution function but differing normalisation. We verify this by normalising each angular slice to the same area under the curve. We observe a very good collapse for all angles with only minor differences for the angle in the original minijet direction. The rescaled distributions agree well with the equilibrium distribution, indicating that even before the isotropisation, at each momentum angle, gluons follow a thermal distribution.

\begin{figure}
    \centering
    \includegraphics{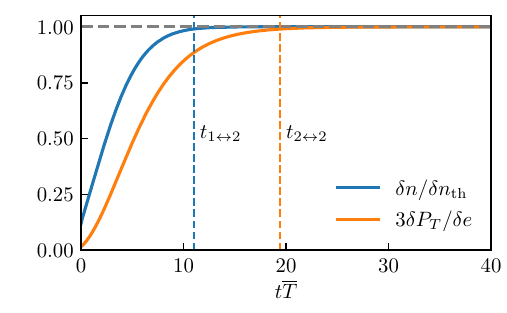}
    \caption{Time evolution of the number density $\delta n$ and the transverse pressure $\delta P_T$, both normalised by the equilibrium value. At the orange and the blue dashed line the difference to $1$ is smaller than $1$\%.}
    \label{fig:PT_n}
\end{figure}

To understand what drives the collapse of distribution functions at different angles \emph{before} the isotropisation, we plot the evolution of particle number $\delta n$ and pressure anisotropy $\delta P_T/\delta e$ in \cref{fig:PT_n}. For a minijet along the $z$ axis, $\delta P_T = \frac{1}{2}(\delta T^{xx}+ \delta T^{yy})$ and $\delta e = \delta T^{tt}$. We see that as the minijet is quenched the particle number $\delta n$ increases. It is driven by the collinear bremsstrahlung of gluons that are emitted while the hard parton interacts with the medium. Similarly, $\delta P_T$ is approaching the equilibrium value thanks to elastic scatterings. However, the density $\delta n$ reaches equilibrium significantly earlier, at time $t_{1\leftrightarrow 2}\overline{T}\approx 11$ (compared to the isotropisation timescale $ t_{2\leftrightarrow 2}\overline{T}\approx 19$). Therefore we conclude that the thermalisation of parity-even perturbations proceeds in two stages: by the time $t_{1\leftrightarrow 2}$ the energy is transported to $p\sim \overline{T}$ and the inelastic processes equilibrate the distribution for each angle separately. Then elastic processes transport energy to larger angles and isotropise the system by $t_{2\leftrightarrow 2}$. It is worth mentioning that this result holds for the whole minijet in \cref{eq:jet_decomposition} since the parity-odd part vanishes upon integrating over the whole phase space.

\subsubsection{Angle-dependent temperature}

\begin{figure}
    \centering
    \includegraphics{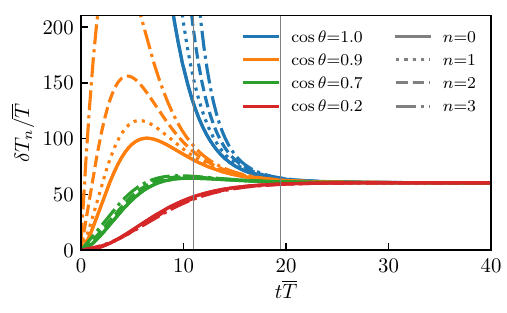}
    \caption{Temperature perturbations $\delta T_{n}$ for different angles $\theta$ (colours) and moments $n$ (linestyles) as a function of time (pure glue). The vertical lines correspond to the two stage thermalisation of the net-energy perturbation in \cref{fig:PT_n}.}
    \label{fig:dI_n_d_L}
\end{figure}

In \cref{fig:df_unscaled_d_E} we showed that over time the distributions collapse to the same functional form, which is actually a thermal distribution. An alternative method of demonstrating this behaviour is to introduce a notion of temperature that is angle-dependent.
 We generalise \cref{eq:Teff} to angular moments of the distribution function $f$ defined by
\begin{equation}\label{I_n}
    I_n(t,\theta) \equiv 4\pi \int\frac{p^2dp}{(2\pi)^3}p^n f(t,p,\theta)= \mathcal{N}_n \times T_n(t,\theta)^{n+3},
\end{equation}
where we multiply with a factor of $4\pi$ instead of integrating over the angles. \Cref{I_n} defines an angle-dependent temperature, which, in the case of a thermal background, can be written as $T_n(\theta) = \overline{T} + \delta T_n(\theta)$.
For linear perturbations, the angle-dependent temperature is given by
\begin{equation}\label{temp_pert}
    \frac{\delta T_n(t,\theta)}{\overline{T}} = \frac{\delta I_n(t,\theta)}{(n+3) \Bar{I}_n},
\end{equation}
where $\delta I_n$ and $\bar I_n$ are the moments of the perturbation and background distributions. We plot the results of \cref{temp_pert} in \cref{fig:dI_n_d_L} for different moments $n$ (solid, dotted, dashed and dot-dashed lines) and different angles $\theta$ (blue, yellow, green, red lines). Around time $t\approx t_{1\leftrightarrow 2}$ the curves collapse for different $n$. Importantly, $\delta T_n(t,\theta)$ is still very different for different angles. From this, it follows that the momentum distribution of the full system $f = \bar{f}_{\text{th}} + \delta f$ along each $\theta$-slice has approximately the shape of a thermal distribution with temperature $T(\theta) = \overline{T} + \delta T(\theta)$. Therefore for times $t>t_{1\leftrightarrow 2}$ the distribution of the minijet is approximately given by a thermal distribution with angle-dependent normalisation (temperature)
\begin{equation}\label{theta_slices_thermal_distr}
    \delta f(t,p,\theta) \approx \delta T(t,\theta) \partial_T f_{\text{th}}(p/T),
\end{equation}
with $\delta T_n(t \gtrsim t_{1\leftrightarrow 2},\theta) \approx \delta T(t,\theta)$ for all $n$. In other words, while the system is still anisotropic, the distribution of the perturbation looks thermal along each angular slice. Then, at $t\approx t_{2\leftrightarrow 2}$, the curves collapse also for each angle $\cos\theta$ and the temperature $\delta T(\theta) = \delta T$ is the same in all directions.

\subsubsection{Jet energy and coupling constant scaling}

Here we study the equilibration time dependence on the initial minijet energy $E$ and coupling constant $\lambda$. In EKT simulations for isotropic under-occupied initial conditions, the thermalisation timescale for high-momentum perturbations was fitted to be~\cite{Kurkela:2014tea}
\begin{equation}
t^\text{under occ.}_\text{eq} \approx  \frac{34.+21. \log(E/T)}{1+0.037 \log \lambda^{-1}} \sqrt{\frac{E}{T}} \frac{1}{\lambda^2 T},
\end{equation}
where $\sqrt{{E}/{T}}/{\lambda^2 T}$ is the parametric timescale for the first hard, medium-induced splitting. As the subsequent splittings proceed faster, the first splitting is an indicative timescale for the energy transfer towards the background energy scale. 
Note that for our initial conditions, the minijet perturbations are anisotropic, and therefore minijet thermalisation is achieved at the isotropisation time $t_{2\leftrightarrow 2}$.

To compare minijet evolution with different initial energies $E$ and different coupling constants $\lambda$ we scale the evolution time with relaxation time
\begin{equation}
t_R=\frac{\eta/s}{\overline{T}} \sqrt{\frac{E}{E_0}},
\end{equation}
where parametrically the specific shear viscosity is $\eta/s\sim \lambda^{-2}$. Then $t_R$ has the same parametric coupling dependence as $t^\text{under occ.}_\text{eq}$. We choose scaling with $\eta/s$ instead of $\lambda$, because it is a physical property of the QGP. We also choose $E_0=30\overline{T}$ as a constant minijet energy reference.
 
\begin{figure}
    \centering
\includegraphics[width=0.49\linewidth]{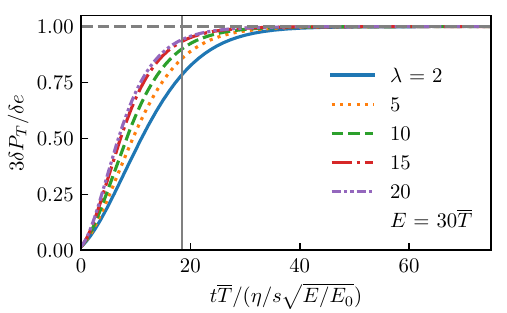}
\includegraphics[width=0.49\linewidth]{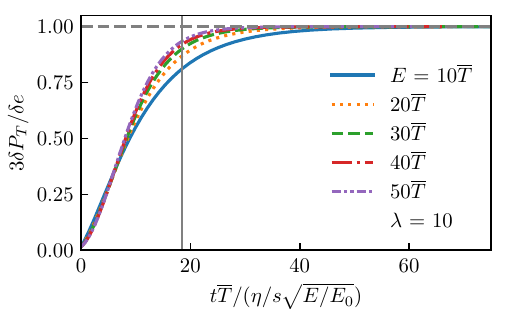}
    \caption{Transverse pressure over energy density as a function of scaled time for (left) different couplings $\lambda$ and (right) different initial energies $E$ of the minijet (pure glue). Gray vertical line indicates $3\delta P_T/\delta e>0.9$ for $E=30 \overline{T}$ and $\lambda=10$.}
    \label{fig:P_T_scaled}
\end{figure}

In \cref{fig:P_T_scaled} we plot the minijet anisotropy $3\delta P_T/\delta e$ as a function of scaled time $t/t_R$ for different coupling strengths (left panel) and initial minijet energies (right panel).
In the left panel, we see similar evolution for different couplings $\lambda=2$, $5$, $10$, $15$, $20$ that correspond to different shear viscosity values ($\eta/s\approx 7.84$, $1.81$, $0.624$, $0.361$, $0.235$) for pure gluon plasma~\cite{Keegan:2015avk}. The isotropisation $3\delta P_T/\delta e>0.9$ is reached at $t\approx 18.5 t_R$ for $E=E_0$ and $\lambda = 10$.
In the right panel of \cref{fig:P_T_scaled} we see that minijet isotropisation proceeds very similarly for minijets of different initial energies $E$ given this time rescaling. 
Therefore the energy dependence of the equilibration time is well captured by the factor of $\sqrt{E}$, in particular for large $E$.

\subsection{Equilibration of net-momentum perturbations}\label{thermal_anti-jet}

\begin{figure}
    \centering
    \includegraphics{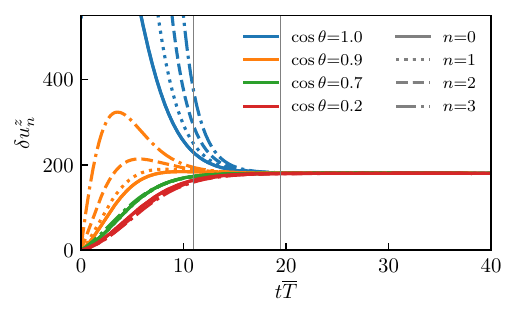}
    \caption{Velocity perturbations $\delta u^z
_{g,n}$ for different angles $\cos\theta$ and moments $n$ as a function of time (pure glue). The vertical lines correspond to the two stage thermalisation observed for net-energy perturbations in \cref{fig:PT_n}.}
    \label{fig:dI_n_a_L}
\end{figure}

Up to now, we have studied parity-even perturbations, which lead to an isotropic increase in the temperature of the system after they thermalise. The parity-odd perturbations, on the contrary, inject net momentum into the system without perturbing the energy density. The evolution of a minijet is a sum of parity odd and even solutions, \cref{eq:jet_decomposition}. We recall from \cref{eq:thjet} that the equilibrium distribution for velocity perturbations contains an explicit $\cos \theta$ factor. Using the moments \cref{I_n} we can again study how the system thermalises in each $\theta$-slice. Analogously to \cref{temp_pert} we now  get
\begin{equation}\label{velo_pert}
    \delta u^z_n(t,\theta)\cos\theta = \frac{\delta I_n(t,\theta)}{(n+3)\Bar{I}_n},
\end{equation}
where in thermal equilibrium $\delta u^z_n(t,\theta)\to \delta u^z$ becomes a velocity field.
In \cref{fig:dI_n_a_L} we show $\delta u^z_n(t,\theta)$ for different moments $n$ and angles $\theta$ as a function of time. 
The collapse of different $n$ moments indicates the emergence of a well-defined velocity field  $\delta u^z(t,\theta)$ at a given angle. However, we notice this happens for all angles at approximately the same time. Therefore there is no significant separation in timescales as there was for temperature perturbations, c.f. \cref{fig:dI_n_d_L}.
We note that the emergence of the velocity field is faster than the temperature field. The reason is that, even in equilibrium, the velocity perturbation is anisotropic due to its $\cos \theta$-dependence. This means that the elastic scatterings do not have to fully isotropise the momentum distribution before reaching the equilibrium distribution.

\subsection{Chemical equilibration}

\begin{figure}
    \centering
    \includegraphics{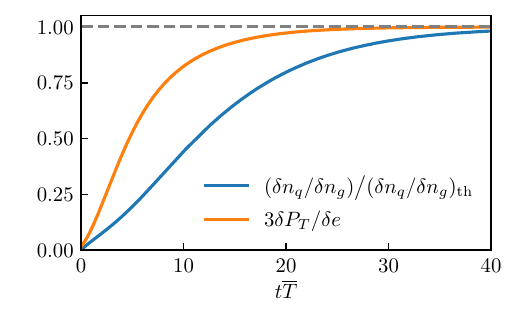}
    \caption{Time evolution of the ratio of quark and gluon number densities $\delta n_q/\delta n_g$ and the transverse pressure $\delta P_T$, both normalised by their equilibrium value.}
    \label{fig:qcd_number_chem_eq.pdf}
\end{figure}

In the previous section, we considered the dynamics of a purely gluonic system. Now we will solve the Boltzmann equation \cref{eq:hom_boltzmann} with
all leading order QCD scattering processes in $C[f]$ for both quark and gluon distributions, $f_q$ and $f_g$. Following~\cite{Mehtar-Tani:2022zwf}, we initiate the evolution with a gluon minijet $\delta f_g$, whereas the initial quark perturbation is set to zero, $\delta f_q(t_0,\p)=0$. This allows us to study the chemical equilibration of minijets. In \cref{fig:qcd_number_chem_eq.pdf} we study the equilibration of quark and gluon number density and total transverse pressure ($P_T=P_T^g+P_T^q$). We observe that the system isotropises first and only at later times the ratio $\delta n_q/\delta n_g$ reaches its equilibrium value.

\begin{figure}
    \centering
    \includegraphics{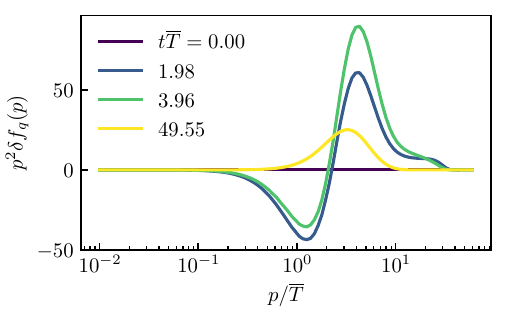}
    \caption{Quark momentum distribution $\delta f_q(\p)$ as a function of time in the minijet direction ($\cos\theta = 1$).}
    \label{fig:q_dfp2_vz_time}
\end{figure}

In \cref{fig:q_dfp2_vz_time}, we show the evolution of the quark distribution along the minijet direction $\cos\theta=1$. Although at late times we expect (and observe) the development of a thermal quark distribution, \cref{eq:thjet}, at the early times  $t\overline{T}\sim 0-4$, the quark perturbation becomes negative at the background scale $p\sim \overline{T}$.
 This can be interpreted as a medium response to the jet. Namely, the hard gluon scatters off of a plasma quark and transfers momentum to the soft particles. Therefore we see a depletion of soft fermions and an increase of higher-momentum quarks. The same process happens for the gluon background, but in this case, there is a competing process of collinear gluon radiation that masks this depletion.

\begin{figure}
    \centering
    \includegraphics[width=0.49\linewidth]{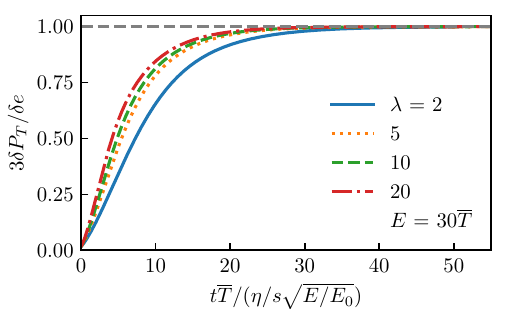}
    \includegraphics[width=0.49\linewidth]{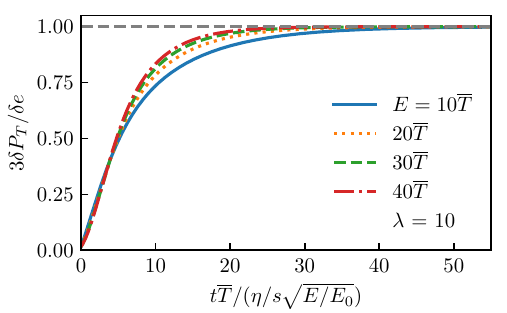}
    \caption{Transverse pressure over energy density $\delta P_T/\delta e$ as a function of scaled time for (left) different couplings $\lambda$ and (right) different initial energies $E$ of the minijet for a QCD plasma including both quarks and gluons.}
    \label{fig:qcd_therma_P_T}
\end{figure}

\begin{figure}
    \centering
    \includegraphics[width=0.49\linewidth]{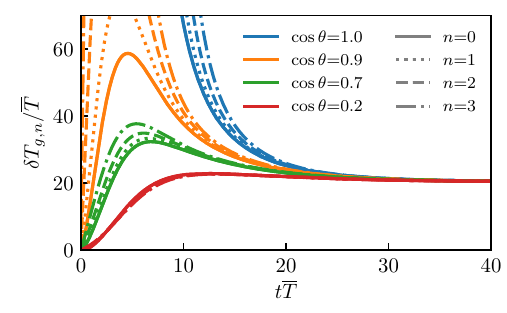}
    \includegraphics[width=0.49\linewidth]{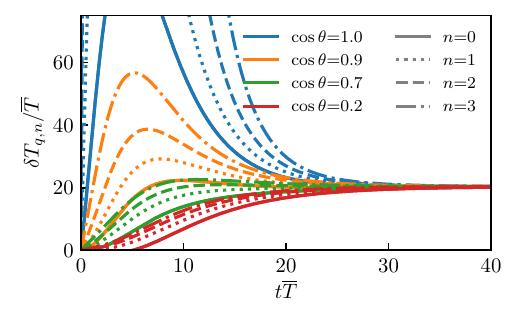}
    \caption{Temperature perturbations $\delta T_{a,n}$ for (left) gluons and (right) quarks as a function of time. Gluons show a very similar behaviour to the pure glue case.}
    \label{fig:qcd_dI_n_d_g_L}
\end{figure}

The isotropisation of the net energy-momentum tensor in the QCD case proceeds in a very similar fashion to YM. As shown in \cref{fig:qcd_therma_P_T}, the coupling constant and minijet energy dependence is very similar. Note, however, in this case the couplings $\lambda=2$, $5$, $10$, $15$, $20$  correspond to  $\eta/s\approx 10$, $2.75$, $0.97$, $0.55$, $0.37$ in a QCD plasma~\cite{Kurkela:2018oqw}. In \cref{fig:qcd_dI_n_d_g_L} we study the temperature evolution extracted from gluon and quark distribution functions. Although the gluons equilibrate kinetically among themselves in a very similar fashion as in the YM case (see also~\cite{Kurkela:2018oqw}), fermion distributions cannot be described by a thermal distribution until the system is chemically equilibrated.


\begin{figure}
    \centering
    \includegraphics[width=0.49\linewidth]{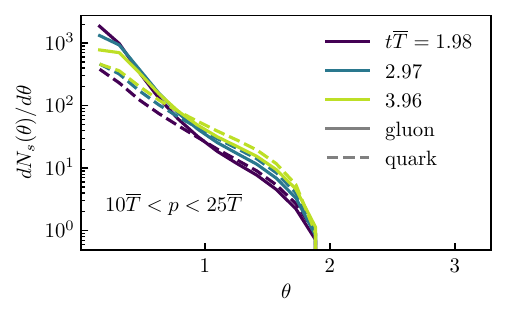}
    \includegraphics[width=0.49\linewidth]{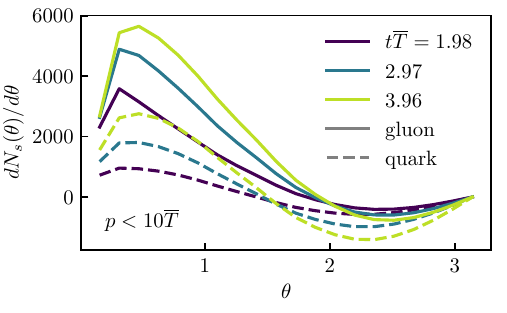}
    \caption{Angular distribution $dN_s(\theta)/d\theta$ of quarks and gluons for (left) semihard and (right) soft particles.}
    \label{fig:dN_dtheta}
\end{figure}

Although in thermal equilibrium quarks are expected to dominate the energy density, already at early times the interaction with the medium changes the chemical composition of the original gluon jet. In Ref.~\cite{Sirimanna:2022zje} this was interpreted as the increase of flavour in quenched jets. In \cref{fig:dN_dtheta} we show the angular composition of high and low momentum partons of a minijet
\begin{equation}
    \frac{dN_s}{d\theta} =\nu_s\sin\theta  \int_{p_\text{min}}^{p_\text{max}} dp \int\frac{d\phi}{(2\pi)^2}p^2\delta f_s(\p),
\end{equation}
where $\delta f_s(\p)$ is the distribution for the original minijet from \cref{eq:jet_decomposition}. Although the kinematics of our simulation is slightly different from  Ref.~\cite{Sirimanna:2022zje}, as we consider lower energy perturbations, we reproduce the qualitative features\footnote{We express time in scaled units of $t\overline{T}$, but physical units of time can be easily recovered by considering a particular background temperature, e.g. $\overline{T}\approx 250\,\text{MeV}=(0.79\,\text{fm})^{-1}$.}. Namely, at high momentum, quarks start to dominate already at early times, while at low momentum gluons are still more abundant. Here we recall from  \cref{fig:q_dfp2_vz_time} that soft quarks below the temperature scale are even depleted initially, which might explain a suppression of the integrated yields below $p<10\,\overline{T}$.

\subsection{Anisotropic background}\label{Anisotropic background}

\begin{figure}
    \centering
    \includegraphics[width=0.49\linewidth]{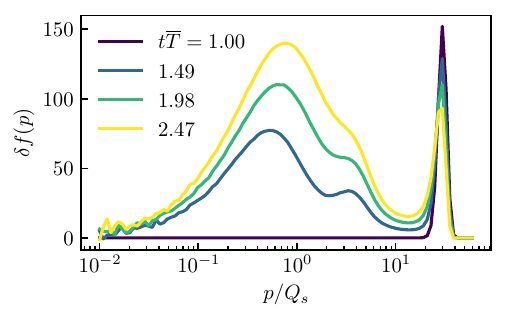}
    \includegraphics[width=0.49\linewidth]{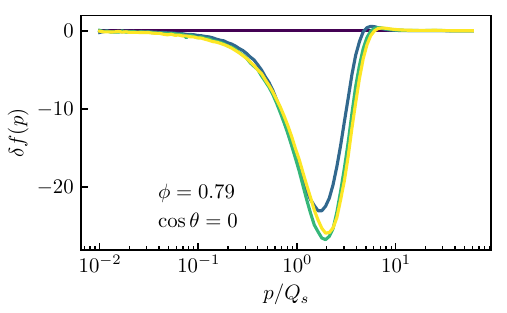}
    \includegraphics[width=0.49\linewidth]{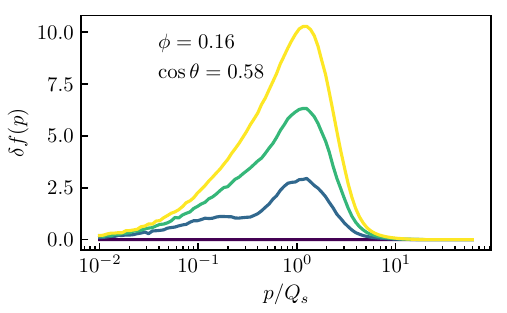}
    \caption{Scattering out of the transverse plane for an anisotropic background. (Top left) Momentum distribution $\delta f(p)$ in minijet direction, $\cos\theta = 0$ and $\phi = 0$, (top right) in the transverse plane, but not in the minijet direction and (bottom) away from the transverse plane.}
    \label{fig:dfp2_phi_time}
\end{figure}

\begin{figure}
\centering
    \includegraphics[width=0.6\columnwidth]{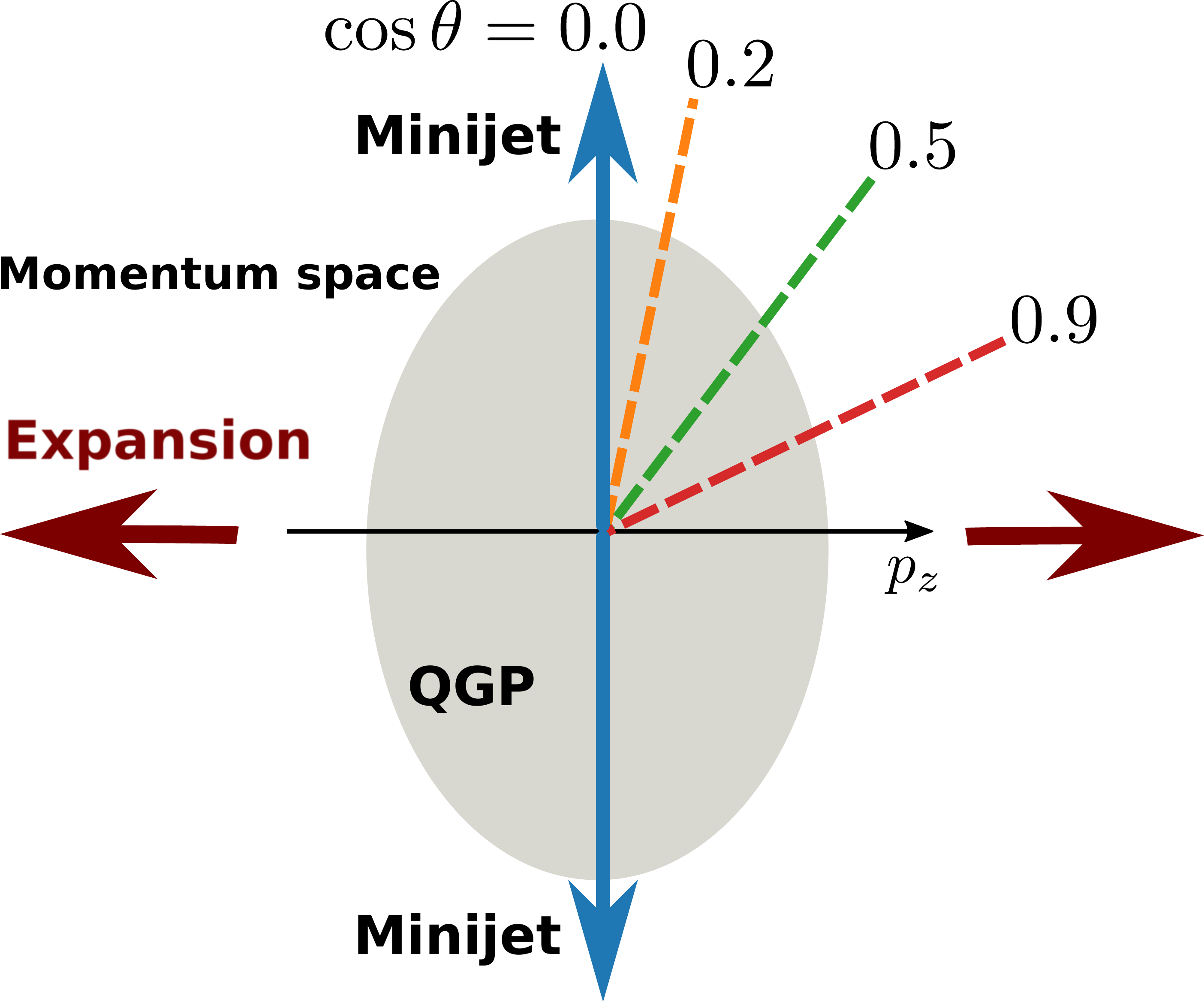}
    \caption{Illustration of a parity-even minijet configuration in momentum space for an out-of-equilibrium QGP that is expanding in $p_z$-direction. For an anisotropic background, we choose the perturbation to point in $x$-direction.\label{fig:dijetcartoon_expanding}}
\end{figure}

Before we move on to discussing the case of an expanding background, we want to address the equilibration of a minijet on top of a non-thermal background without expansion. Even though the system is not expanding, the $z$-direction is singled out in the initial distribution of gluons, \cref{init_prl_distr}, since we choose $\xi \gg 1$. This means that gluons in the $z$ direction are highly suppressed, while in the transverse plane, they are highly occupied. We choose the initial perturbation $\delta f_g(t_0,\p)$ to be in the transverse plane and pointing in $x$-direction, i.e., $\cos\theta=0$, $\phi=0$ (cf. \cref{fig:dijetcartoon_expanding}). Hence, the loss term in the elastic kernel $C^g_{2\leftrightarrow 2}(\p_{\perp},p_z=0)$ receives large contributions compared to the gain term, since the incoming particles are suppressed for non-zero $p_z$. In other words, due to the shape of the background $\Bar{f}_g \left(\tau_0,p,\theta\right)$, the initial minijet perturbation is scattered out of the plane, while the scatterings into the transverse plane are suppressed. Note that in this case the cubic $\sim f^3$ terms in the elastic collision kernel are suppressed for scatterings into an empty region of the phase space and the subleading $\sim f^2$ terms are responsible for the onset of isotropisation~\cite{Epelbaum:2015vxa}.
As a consequence, for early times, the minijet perturbation becomes negative for large $\phi$. This can be seen in \cref{fig:dfp2_phi_time}, where we compare the evolution of the momentum distribution $\delta f(p)$ in different directions. Close to the jet-axis (top panel), this effect is covered up by the collinear splittings, but one can see a dip around $p \sim 2 Q_s$. Away from the minijet axis, we have $\delta f<0$ in the transverse plane (middle panel) and $\delta f>0$ away from the transverse plane indicating out-of-plane scattering.

\begin{figure}
    \centering
    \includegraphics{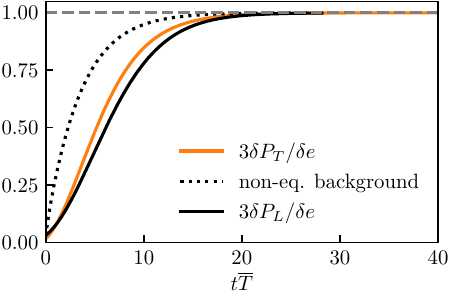}
    \caption{The time evolution of the minijet pressure perpendicular to the minijet direction in thermal and anisotropic background normalised to the equilibrium value. The dotted line shows the isotropisation of the anisotropic background.}
    \label{fig:both_PT_with_bg}
\end{figure}

In \cref{fig:both_PT_with_bg} we follow the pressure anisotropy evolution of the background (black dotted line) and the mini-jet perturbation (solid black line)\footnote{In our coordinate system the initial minijets point in perpendicular directions in the anisotropic case compared to the thermal one, therefore the pressure \emph{perpendicular} to the minijet direction changes from $P_{T}$ to $P_{L}$, see Fig.~1 and Fig.~13.}. To compare to the thermal background results in Fig.~3 we rescale the time using the effective temperature $\overline{T}$, which is obtained from Landau matching to the energy density, Eq.~(2.14). Without expansion, the energy density is conserved and $\overline{T}$ corresponds to the final temperature of the background. We see that, somewhat surprisingly, perturbations on isotropic (orange) and non-anisotropic (black) backgrounds follow very close trajectories towards isotropy. Because the background distribution isotropise at an earlier time, the initial background anisotropy does not appear to significantly delay the equilibration of the mini-jet.

\section{Minijet hydrodynamisation in expanding plasma}\label{exp_plasma}

In this section, we study minijet equilibration in a plasma undergoing Bjorken expansion. Namely, we solve \cref{eq:linearised1,eq:linearised2} for quark and gluon distributions, where the time variable is the proper time $\tau=\sqrt{t^2-z^2}$. As in \cref{Anisotropic background}, we now consider a minijet perturbation on top of an anisotropic background parametrised by \cref{init_prl_distr}. These CGC-inspired initial conditions are highly occupied but anisotropic, and have been used previously to describe bulk QGP hydrodynamisation and chemical equilibration~\cite{Kurkela:2015qoa,Kurkela:2018oqw,Kurkela:2018xxd,Kurkela:2018vqr}. These studies have shown that for moderate couplings the evolution becomes universal if expressed in units of the time-dependent relaxation time $\tau_R(\tau)=(4\pi\eta/s)/\overline{T}(\tau)$, i.e., in terms of the \emph{scaled time} $\Tilde{w} = \tau/\tau_R$.

At late times the particle distributions approach the equilibrium distribution, \cref{therm_distr}, with time-dependent temperature $\overline{T}\propto \tau^{-1/3}$. In addition, the asymmetric plasma expansion $\partial_\mu u^\mu=1/\tau$, deforms the distribution by viscous corrections proportional to these gradients and the shear viscosity $\eta$. This is encoded in the shear stress tensor $\pi^{\mu\nu}$ which describes the deviation of the energy-momentum tensor from ideal hydrodynamics. Generally, we have~\cite{Teaney:2003kp}
\begin{equation}
    \delta f_\text{visc} \propto p_\mu p_\nu \pi^{\mu\nu}.
\end{equation}
In a Bjorken expanding system, $\pi^{\mu\nu}$ is a diagonal matrix leading to the general form
\begin{equation}
\delta f_\text{visc} = \frac{\eta/s}{\tau \overline{T}(\tau)}(1-3\cos^2 \theta)F\left(p/\overline{T}(\tau)\right),
\end{equation}
of linearised viscous corrections. $F(p/\overline{T})$ is some isotropic function of momentum that is not known analytically but can be extracted from numerical simulations~\cite{Dusling:2009df}. Viscous corrections decay only as a power of time, whereas we will see that minijets are quenched on shorter timescales. In the following, we will study how minijet perturbations equilibrate in such a background and how the separation of the initial minijet energy $E$ from the background energy scale $Q_s$, and the direction of the jet, affect the equilibration of the jet.

\subsection{Hydrodynamisation}
\subsubsection{Net-energy perturbations}\label{exp_di-jet}

While the minijet perturbation equilibrates, the background $\Bar{f}(\tau_0,\p)$ undergoes the thermalisation process as well. After being highly occupied initially, it approaches a form that is well described by viscous hydrodynamics~\cite{Kurkela:2015qoa}. For that reason, we are interested in the timescales at which the minijets become part of the hydrodynamic background. However, in contrast to the non-expanding case, we do not have an analytical expression for the equilibrated minijet perturbation as in \cref{eq:thjet}. Instead, we will use the fact that the system loses its memory of the initial conditions once the minijet hydrodynamises. In addition to jet-like perturbations, we will also consider the kinetic evolution of perturbations that are very different, and consider the perturbations to have hydrodynamised once the system evolution is the same for both.
To this end, we use the background distribution $\Bar{f}(\tau_0,\p)$ as a perturbation
\begin{equation}
    \delta f^{\text{az}}_{\text{sym}}(\tau_0,\p) = \frac{\delta A}{A}f_\text{sat}(p_T, p_z).\label{eq:azymdf}
\end{equation}
In contrast to the types of perturbations that may describe a jet, this perturbation is azimuthally symmetric.

\begin{figure}
    \centering
    \includegraphics{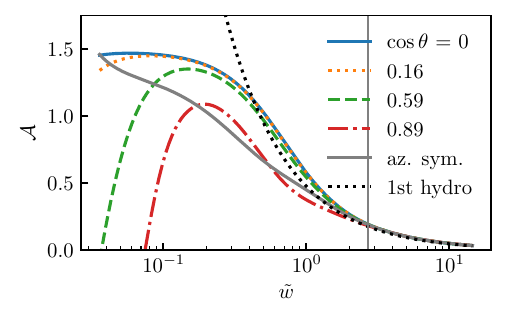}
    \caption{Pressure anisotropy $\mathcal{A}$ for different initial orientations $\cos\theta$ of the minijet as a function of time. Once all the curves are in a range of $1\%$ within each other, we say that they have collapsed (vertical line). The black dotted line is the expected evolution from viscous hydrodynamics.}
    \label{fig:exp_qcd_anis_angles}
\end{figure}

A common quantity used to study hydrodynamisation is the pressure anisotropy~\cite{Florkowski:2017olj}
\begin{equation}
    \mathcal{A} = \frac{\delta P_T - \delta P_L}{\delta e/3}.
\end{equation}
Close to equilibrium the hydrodynamic prediction is $\mathcal{A} = \frac{3}{2\pi}\tilde{w}^{-1}$. In \cref{fig:exp_qcd_anis_angles} we plot the anisotropy $\mathcal{A}(\tilde{w})$ for jets at different angles with the expansion axis, as well as the azimuthally symmetric perturbation. 
For a minijet in the transverse plane ($\cos\theta=0$), the longitudinal pressure vanishes and $\mathcal{A}=3/2$, while jets in the $z$-direction start with positive $\delta P_L$ and $\mathcal{A}<0$. However, the longitudinal pressure is rapidly suppressed by expansion and $\mathcal{A}$ becomes positive. By the time of $\tilde w_\text{mjh}\approx2.7$ jets with different directions from the expansion axis, and the azimuthally-symmetric perturbation, have the same asymmetry as one another (and as the hydrodynamic prediction) within 1\%. We call it the minijet hydrodynamisation time. 

\begin{figure}
    \centering
    \includegraphics{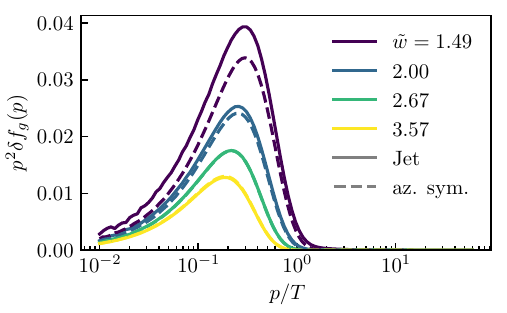}
    \caption{Normalised gluon distributions $ \delta f_g$  of the minijet and the azimuthally symmetric distribution $\delta f^{\text{az}}_{\text{sym}}$, in minijet direction ($\phi=0$ and $\cos\theta=0$).}
    \label{fig:exp_df_ref}
\end{figure}
In \cref{fig:exp_df_ref} we explicitly compare the gluon distribution function of the jet-like perturbation in the transverse plane and the azimuthally-symmetric perturbation at late times. Initially, the minijet points in the $x$-direction ($\phi=0$ and $\cos\theta=0$). By the time of $\tilde w_\text{mjh}\approx 2.7$ the distributions agree with each other.

\begin{figure}
    \centering
    \includegraphics[width=0.49\linewidth]{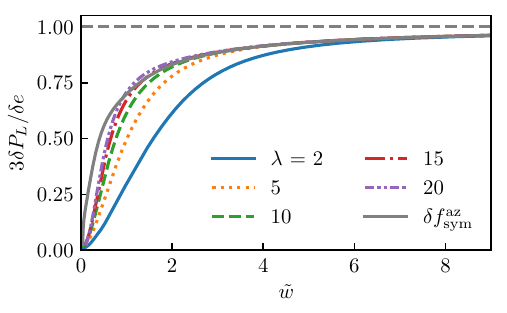}
    \includegraphics[width=0.49\linewidth]{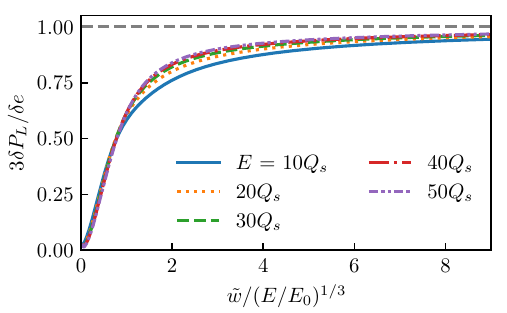}
    \caption{Longitudinal pressure over energy $\delta P_L/\delta e$ as function of time for (left) different couplings $\lambda$ and (right) different energies $E$. In the top panel we compare the evolution with the azimuthally symmetric perturbation $\delta f^{\text{az}}_{\text{sym}}$.}
    \label{fig:exp_P_L_scaled_L2}
\end{figure}

Next, we study how minijet evolution depends on their initial energy and the coupling strength. The left panel of \cref{fig:exp_P_L_scaled_L2} shows that the scaled time $\tilde w$ captures well the coupling dependence, especially for larger values of the coupling constant. In particular, by $\tilde w_\text{mjh}=2.7$ all simulations with coupling $\lambda\geq 5$ follow the same trajectory.
In the right panel of \cref{fig:exp_P_L_scaled_L2} we see that the minijet energy dependence is also well captured by  $\sqrt[3]{E}$ dependence. The cubic root raises from the non-linear time dependence of $\tilde w \propto \tau^{2/3}$ close to equilibrium. 
We conclude that for sufficiently large couplings and initial energies, all minijets hydrodynamise by the time $\tilde w_\text{mjh} = 2.7 \left({E}/{E_0}\right)^{1/3}$. 

We can convert the rescaled time in dimensionful units following the arguments of~\cite{Kurkela:2018wud} and solving
\begin{align}
\frac{\tau_\text{mjh} T(\tau_\text{mjh})}{4\pi\eta/s}  = 2.7 \left(\frac{E}{E_0}\right)^{1/3} \label{eq:X}
\end{align}
for the minijet hydrodynamisation time $\tau_\text{mjh}$.
At late times, the temperature evolution approaches the ideal hydro prediction 
\begin{equation}
T(\tau) = \frac{\Lambda_T}{(\Lambda_T \tau)^{1/3}},\label{eq:Tid}
\end{equation}
where $\Lambda_T$ is an asymptotic energy scale, which we can determine from the average entropy per unity rapidity in hydrodynamic simulations
\begin{equation}
\left<s\tau\right> = \nu_\text{eff}\frac{4\pi^2}{90} \Lambda_T^2.
\end{equation}
Thanks to near-ideal hydrodynamic evolution, it is proportional to the produced particle multiplicity and is, therefore, constrained by data~\cite{Hanus:2019fnc}.

Substituting \cref{eq:Tid} into \cref{eq:X} we can solve for $\tau_\text{mjh}$
\begin{align}
\tau_\text{mjh} &= 2.7^{3/2}\frac{(4\pi \eta/s)^{3/2}}{\Lambda_T}\left(\frac{E}{E_0}\right)^{1/2}
\end{align}
The reference energy is  $E_0=30Q_s$, which can be expressed in terms of $\Lambda_T=0.47 Q_s$ for the background QGP equilibration for $\lambda=10$. 
Then choosing the typical values of physical parameters $\eta/s\approx 0.16$, $\left<s\tau\right>=4.1\,\text{GeV}^2$ and $\nu_\text{eff}\approx 40$~\cite{Kurkela:2018wud} we obtain

\begin{align}
\tau_\text{mjh} &= 5.1\,\text{fm}\left(\frac{4\pi\eta/s}{2}\right)^{3/2}\left(\frac{\left<s\tau\right>/\nu_\text{eff}}{4.1\,\text{GeV}^2/40}\right)^{-3/4} \left( \frac{E}{31\,\text{GeV}}\right)^{1/2}\label{eq:taujh}
\end{align}
This timescale is much larger than the typical background hydrodynamisation time of $\tau_\text{hydro}^\text{QGP}\sim 1.1\,\text{fm}$~\cite{Kurkela:2018wud} and comparable to the total QGP lifetime in central nucleus-nucleus collisions~\cite{Hanus:2019fnc}. 

From \cref{eq:taujh} one would conclude that $E\approx 31\,\text{GeV}$ partons should be fully quenched in a collision. 
However, one should keep in mind that this estimate applies to a parton that is approximately on-shell and only undergoing medium-induced radiation. A $31\,\text{GeV}$ parton produced in a hard scattering would have virtuality and therefore undergo vacuum-like fragmentation in addition to the processes considered here. Fragmentation breaks a high-energy parton into several lower energy ones on timescales faster than those for typical medium interactions. Therefore we anticipate that it would further decrease the quenching time compared to the estimate in \cref{eq:taujh} for on-shell parton. 
Measurements of the jet nuclear modification factor at this energy, for example~\cite{ALICE:2019qyj}, show that jets of such energy are significantly suppressed compared to the proton collisions. However, the degree of quenching can be seen more directly in measurements of the momentum correlations between a direct photon and the recoiling jet~\cite{CMS:2017ehl,ATLAS:2018dgb}. Though these measurements are for slightly higher momenta, most jets retain a  substantial fraction of their energy. Therefore the quenching time estimate from \cref{eq:taujh} seems too high.

One possible reason for short hydrodynamisation times is that in our simulations the interaction strength is controlled by $\eta/s$ for both the background and perturbations since there is no running of the coupling constant at leading order. In contrast, current energy loss models use a separate parameter, $\hat{q}/T^3$, to set the strength of jet quenching. In reality, it is likely that high-momentum partons will be more weakly coupled to the medium than medium-medium interactions which would tend to decrease the degree of quenching.
Though presumably less important than the previously-mentioned effects, our estimate is also based on boost-invariant simulations, which becomes less accurate when the QGP transitions to 3D expansion at times comparable to the typical system size. 
Therefore the validity of \cref{eq:taujh} improves for smaller initial parton energies, where the effect of running coupling and the 3D expansions is smaller.

\subsubsection{Net-momentum perturbations}

\begin{figure}
    \centering
    \includegraphics{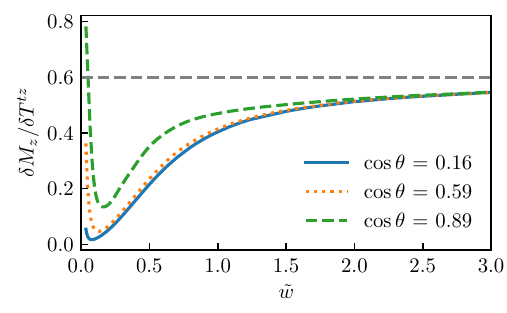}
    \caption{The moments $\delta M_z$ as a function of time, normalised by the flux $\delta T^{tz}$. Different lines correspond to different initial orientations of the minijet with respect to the expansion axis.}
    \label{fig:exp_qcd_Iz_T0z}
\end{figure}

Analogously to the static case studied in  \cref{thermal_anti-jet}, we would like to investigate the equilibration of the parity-odd perturbations in the expanding background.
These parity-odd perturbations add to the momentum of the system, $\delta T^{ti}$, but keep the other diagonal components of the energy-momentum tensor, like energy density $\delta e$ or the pressure $\delta P_{T,L}$, unchanged. Therefore to monitor its equilibration, we need to introduce new parity-odd moments of $\delta f$. To this end, we define
\begin{equation}
    \delta M_z = \sum_a \nu_a \int \frac{d^3\p}{(2\pi)^3}\frac{(p_z)^3}{p^2}\delta f_a(\p),
\end{equation}
where $p_z/p = \cos\theta$ and $\theta$ is the angle to the beam axis. It is straightforward to show that in thermal equilibrium we have that $\delta M_z = \frac{3}{5}\delta T^{tz}$ (see \cref{eq:thjet}).

In \cref{fig:exp_qcd_Iz_T0z} we plot the time evolution of the ratio $\delta M_z/\delta T^{tz}$ as a function of $\Tilde{w}$ for different initial angles between the perturbation and the expansion axis. Following a strong decrease due to the rapid expansion in the beginning, $\delta M_z$ approaches the equilibrium value for all angles between the minijet and expansion axes.  The evolution for different angles $\cos\theta$ is the same beginning around $\Tilde{w}\approx 2$, which we interpret as hydrodynamisation time for parity-odd perturbations. This time is somewhat smaller than for the parity even perturbations, cf. \cref{fig:exp_qcd_anis_angles}. This is consistent with the finding in \cref{thermal_anti-jet} that for a static background, the velocity field $\delta u^z$ is built up faster than the temperature field $\delta T$.

\subsection{Chemical equilibration}
\begin{figure}
    \centering
    \includegraphics{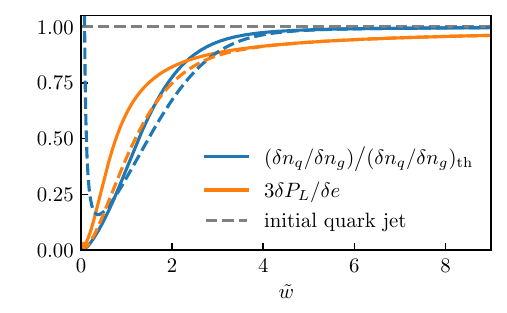}
    \caption{Time evolution of the ratio of quark and gluon number densities $\delta n_q/\delta n_g$ and the transverse pressure $\delta P_T$ in an expanding system. Both quantities are normalised by their equilibrium value. Solid (dashed) lines correspond to the evolution of an initial gluon (quark) jet.}
    \label{fig:exp_qcd_number_chem_eq}
\end{figure}

We will now study how the quark and gluon fraction of the minijets evolve within an expanding background. In \cref{fig:exp_qcd_number_chem_eq} we show the time evolution of the ratio of quark and gluon densities $\delta n_q/\delta n_g$ compared to the build-up of longitudinal pressure $\delta P_L$, as in \cref{fig:qcd_number_chem_eq.pdf}. Interestingly, the order of chemical and kinetic equilibration is now reversed. This is consistent with the ordering found for the background evolution in Bjorken expansion~\cite{Kurkela:2018oqw,Kurkela:2018xxd}. The longitudinal expansion generates pressure anisotropy proportional to the expansion rate $\partial_\mu u^\mu=1/\tau$ and specific shear viscosity $\eta/s$. However, expansion does not imbalance the chemical composition of the plasma and chemical equilibration can be achieved faster than isotropisation.

So far, we have studied minijets initiated by a perturbation in the gluon distribution. However, we can similarly study the evolution of a non-zero quark perturbation $\delta f_q(\tau_0,\p)$ and set the initial gluon distribution to zero. The evolution for such an initial condition is depicted with dashed lines in \cref{fig:exp_qcd_number_chem_eq}.  We observe that gluons are produced rapidly, and the perturbation becomes gluon-dominated at $\tilde{w}<1$. From then on, the evolution is qualitatively similar to the case of an initial gluon jet, although the equilibration is slightly delayed. The pressure $P_L$ takes more time to isotropise in the beginning, but by around $\tilde{w}_\text{mjh}\approx2.7$ both quark- and gluon-initiated perturbations have reached the same anisotropy.

\section{Conclusions}\label{sec:concl}

In this paper, we used a leading-order QCD effective kinetic theory to study the thermalisation of high-momentum low-virtuality partons (minijets) in the QGP. Using a unifying framework of kinetic theory we were able to simultaneously describe the evolution of the background QGP and the thermalisation of a linearised high-momentum perturbation.  Consequently, both the high-momentum energy loss and low-momentum transport are governed by the same parameter---the QCD coupling constant---which we quantify by the physical property of QGP, the specific shear viscosity $\eta/s$.
Improving on the previous work of minijet quenching in thermal backgrounds, we studied minijet thermalisation and hydrodynamisation in a Bjorken-expanding QGP.

In the first part of the paper, we studied the equilibration of minijets in a static thermal background and found that the two conserved quantities (energy and momentum) are deposited and thermalise in slightly different ways. Energy is deposited via a two-step cascade, where the energy of the high-momentum parton first flows to the soft sector by collinear cascade and then is isotropised by elastic scatterings. This results in a scaling of the distribution function characterised by an angle-dependent temperature, prior to isotropisation. 
In contrast, the momentum deposition to the thermal plasma proceeds faster and we observe no two-step thermalisation. 
We find that pressure isotropisation precedes chemical equilibration for non-expanding plasma, which is the same as for the background QGP~\cite{Kurkela:2018oqw,Kurkela:2018xxd}.

In the second part of the paper, we studied minijets embedded in an expanding and non-equilibrium QGP. Because the longitudinal expansion breaks the rotational symmetry, we considered minijets initialised at different angles with respect to the direction of the expansion. We found that the longitudinal expansion rapidly inverses the pressure anisotropy of minijets pointing towards the beam axis. Performing simulations for different coupling values and initial minijet energies, we find that high-momentum perturbations hydrodynamise, meaning that they become indistinguishable from soft background perturbations, at the rescaled time
$\tilde w_\text{mjh} \approx 2.7\sqrt[3]{E/E_0}$, where $\tilde w=\tau T(\tau)/(4\pi\eta/s)$. 
Although we mostly studied gluon-initiated perturbations, we also found that quark-initiated jets are quickly populated by gluons and the subsequent evolution is qualitatively the same as for gluon jets.

We presented a detailed study of how high-momentum perturbations, minijets, are quenched by the QCD medium in and out-of-equilibrium in leading order QCD kinetic theory. However, our study involved several simplifications that should be relaxed for a more realistic description of minijets in heavy-ion collisions. First, we considered perturbations that are homogeneous in space and boost-invariant. Therefore our results could be thought of as spatially-averaged but momentum-resolved evolution of minijets. 
Similarly, at late times the transverse dynamics of the QGP can no longer be neglected. Therefore relaxing the assumptions about homogeneity and boost-invariance of both background and perturbations is an important next step in the detailed studies of jet medium-interactions. However, such simulations are computationally demanding as the dimensionality of the phase-space increases. Second, we use leading order QCD kinetic theory and neglect the running coupling effects. Thermalisation dynamics of the QGP at next-to-leading order has been studied for isotropic systems in Ref.~\cite{Fu:2021jhl}. The jet-medium interactions at NLO in isotropic QGP are known but have not been studied numerically~\cite{Ghiglieri:2015ala}. It would be interesting to study the effect of higher-order corrections on minijet quenching. However, it will be difficult to describe the final stages of minijet thermalisation, where the full control of NLO corrections in an anisotropic medium is needed~\cite{Ghiglieri:2018dib}.

Although QCD kinetic theory provides a dynamical non-equilibrium picture of QGP, it is based on a series of approximations of the underlying quantum field theory~\cite{Berges:2020fwq, Arnold:2002zm}. First, the de Broglie wavelength of the quasiparticles must be small compared to the mean free path between collisions and quantum interference effects between successive scattering events are neglected. These conditions are satisfied in a weakly coupled QGP~\cite{Arnold:2002zm}. In addition, the Boltzmann equation is a leading order gradient expansion of the full quantum evolution~\cite{Berges:2004pu}. In kinetic theories with non-local interactions~\cite{Weickgenannt:2021cuo,Sheng:2021kfc}, there could be higher order gradient corrections in the Boltzmann equation~\cite{Barata:2022utc}. Furthermore, the QCD kinetic theory describes radiation rates as those from an infinite medium. Therefore the finite-size effects that are taken into account in the BDMPS-Z type computations are neglected in kinetic rates. In particular, newly produced partons are resolved instantaneously, leading to higher radiation rates than in BDMPS-Z~\cite{Caron-Huot:2010qjx}. This potentially contributes to rather short minijet quenching times in our work. Taking into account the finite-size effects in kinetic theory simulations is non-trivial,  but would improve the description of energy loss in small collision systems.

The minijet partons that we have studied in this work should be interpreted as the by-products of a jet vacuum shower, and therefore, our work contributes to a better understanding of jet-medium interactions. This work represents a key step towards future developments aimed at incorporating QCD kinetic theory treatments of equilibration into jet phenomenology. 
In the future, our results could be used to describe the thermalisation of jets as a composition of the thermalisation of several minijets, using a factorised vacuum-medium approach such as in Refs.~\cite{Caucal:2018dla,Caucal:2019uvr}.
A promising avenue is to use QCD kinetic theory simulations to extract medium response functions to minijet perturbations. Such an approach is taken in the K\o{}MP\o{}ST framework~\cite{Kurkela:2018vqr,Kurkela:2018wud} for initial net-energy and net-momentum perturbations at the typical medium energy scale. The non-equilibrium K\o{}MP\o{}ST response functions then describe how spatial fluctuations of energy-momentum tensor in initial state hydrodynamise and contribute to the subsequent hydrodynamic evolution. Analogously, one should be able to construct suitable response functions to initial minijet perturbations. 
Another interesting avenue is to use the QCD kinetic theory simulations to provide the dynamical thermalisation picture in the initial state models based on perturbative minijet production~\cite{Niemi:2015qia,Garcia-Montero:2023gex}.

We leave these exciting and challenging undertakings to ongoing and future work.

\begin{acknowledgments}
We thank Jussi Auvinen, Jürgen Berges, Kari Eskola, Jacopo Ghiglieri, Sigtryggur Hauksson, Yuuka Kanakubo, Aleksi Kurkela, Florian Lindenbauer, Harri Niemi, Sören Schlichting, Adam Takacs and Derek Teaney for useful discussions. We are grateful to Jan Philipp Klinger for carefully reading the manuscript. AM and FZ acknowledge support by the DFG through Emmy Noether Programme (project number 496831614)and CRC 1225 ISOQUANT (project number 27381115). JB acknowledges support through the Leverhulme-Peierls fellowship funded by the generosity of the Leverhulme Trust. The authors acknowledge support by the state of Baden-Württemberg through bwHPC and the German Research Foundation (DFG) through grant no INST 39/963-1 FUGG (bwForCluster NEMO).
\end{acknowledgments}

\bibliographystyle{JHEP}
\bibliography{apssamp}

\clearpage

\appendix

\section{Linearised collision kernels}\label{lin_kernels}

The linearised collision kernel $\delta C[\bar{f},\delta f]$ consists of elastic $2\leftrightarrow 2$ scatterings and inelastic effective $1\leftrightarrow 2$ splitting and merging processes. Once we linearise, we get one contribution from the gain and loss terms and one contribution from the matrix elements.

\subsection{$2\leftrightarrow 2$-processes}

First we consider the $2\leftrightarrow 2$-processes. The unperturbed collision term can be written as
\begin{equation}
\begin{aligned}
C^s_{2\leftrightarrow 2}[f](\ptilde) &= \frac{1}{2}
\frac{1}{\nu_s} \frac{1}{4}
 \sum_{abcd} (2\pi)^3 \int_{\p\k\p'\k'}
 |\mathcal{M}^{ab}_{cd} 
 |^2(2\pi)^4
 \delta^{(4)}(P+K-P'-K')\\
&\times \{ (f^a_{\p} f^b_{\k} (1\pm f^c_{\p'})(1\pm f^d_{\k'}))-(f^c_{\p'} 
f^d_{\k'} (1\pm f^a_{\p})(1 \pm f^b_{\k})) \}\\
&\times 
\left[\delta(\ptilde-\p)\delta_{as}+\delta(\ptilde-\k)\delta_{bs} 
-\delta(\ptilde-\p')\delta_{cs}-\delta(\ptilde-\k')\delta_{ds}\right],
\end{aligned}
\end{equation}
The linearised kernel can be split into two parts
\begin{equation}
    \delta C^{s}_{2\leftrightarrow 2}[f](\ptilde) = \delta C^{s,(1)}_{2\leftrightarrow 2}[f](\ptilde) + \delta C^{s,(2)}_{2\leftrightarrow 2}[f](\ptilde),
\end{equation}
and first we will look at the linearisation of the gain and loss terms $\delta C^{s,(1)}_{2\leftrightarrow 2}[f](\ptilde)$. This can be straightforwardly carried out and we get
\begin{equation}
\begin{aligned}
\delta C^{s,(1)}_{2\leftrightarrow 2}[f](\ptilde) &= \frac{1}{8\nu_{s}} (2\pi)^3\int_{\p\k\p'\k'} \Bigg\{\Big[ |\mathcal{M}^{gg}_{gg }|^2 \Big](\delta f^g_{\p} f^g_{\k} (1+f^g_{\pp}) (1+f^g_{\kp})- \delta f^g_{\k} (1+f^g_{\k}) f^g_{\pp} f^g_{\kp} \\
 &+f^g_{\p} \delta f^g_{\k} (1+f^g_{\pp}) (1+f^g_{\kp})- (1+f^g_{\k}) \delta f^g_{\k} f^g_{\pp} f^g_{\kp} \\
 &+f^g_{\p} f^g_{\k} \delta f^g_{\pp} (1+f^g_{\kp})- (1+f^g_{\k}) (1+f^g_{\k}) \delta f^g_{\pp} f^g_{\kp}\\
 &+f^g_{\p} f^g_{\k} (1+f^g_{\pp}) \delta f^g_{\kp}- (1+f^g_{\k}) (1+f^g_{\k}) f^g_{\pp} \delta f^g_{\kp})\\
 &\times \left[ \delta(\tilde{{\p}}-{\p})\delta_{gs}+\delta(\tilde{\p}-\k)\delta_{gs}-\delta(\tilde{\p}-\pp)\delta_{gs}-\delta(\tilde{\p}-\kp)\delta_{g s}\right] \\
&+ \Big[2    |\mathcal{M}^{gg}_{q_1\bar{q}_1}|^2\Big] (\delta f^g_{\p} f^g_{\k} (1-f^q_{\pp}) (1-f^q_{\kp})- \delta f^g_{\k} (1+f^g_{\k}) f^q_{\pp} f^q_{\kp} \\
 &+f^g_{\p} \delta f^g_{\k} (1-f^q_{\pp}) (1-f^q_{\kp})- (1+f^g_{\k}) \delta f^g_{\k} f^q_{\pp} f^q_{\kp} \\
 &+f^g_{\p} f^g_{\k} (-\delta f^q_{\pp}) (1-f^q_{\kp})- (1+f^g_{\k}) (1+f^g_{\k}) \delta f^q_{\pp} f^q_{\kp}\\
 &+f^g_{\p} f^g_{\k} (1-f^q_{\pp}) (-\delta f^q_{\kp})- (1+f^g_{\k}) (1+f^g_{\k}) f^q_{\pp} \delta f^q_{\kp} )\\
 &\times \left[ (2 N_f) \delta(\tilde{\p}-\p)\delta_{gs}+(2 N_f) \delta(\tilde{\p}-\k)\delta_{gs}-\delta(\tilde{\p}-\pp)\delta_{qs}-\delta(\tilde{\p}-\kp)\delta_{qs}\right] \\
&+\Big[4  |\mathcal{M}^{q_1 g}_{q_1 g}|^2\Big](\delta f^q_{\p} f^g_{\k} (1-f^q_{\pp}) (1+f^g_{\kp})- (-\delta f^q_{\p}) (1+f^g_{\k}) f^q_{\pp} f^g_{\kp} \\
 &+f^q_{\p} \delta f^g_{\k} (1-f^q_{\pp}) (1+f^g_{\kp})- (1-f^q_{\p}) \delta f^g_{\k} f^q_{\pp} f^g_{\kp} \\
 &+f^q_{\p} f^g_{\k} (-\delta f^q_{\pp}) (1+f^g_{\kp})- (1-f^q_{\p}) (1+f^g_{\k}) \delta f^q_{\pp} f^g_{\kp}\\
 &+f^q_{\p} f^g_{\k} (1-f^q_{\pp}) \delta f^g_{\kp})- (1-f^q_{\p}) (1+f^g_{\k}) f^q_{\pp} \delta f^g_{\kp} )\\ 
 &\times \left[ \delta(\tilde{\p}-\p)\delta_{qs}+(2 N_f)\delta(\tilde{\p}-\k)\delta_{gs}-\delta(\tilde{\p}-\pp)\delta_{qs}-(2 N_f)\delta(\tilde{\p}-\kp)\delta_{g s}\right] \\
&+\Big[  2[2 (N_f-1)] |\mathcal{M}^{q_1 q_2}_{q_1 q_2}|^2+[2 (N_f-1)] |\mathcal{M}^{q_1 \bar{q}_1}_{q_2\bar{q}_2}|^2+2  |\mathcal{M}^{q_1 \bar{q}_1}_{q_1\bar{q}_1}|^2+ |\mathcal{M}^{q_1q_1}_{q_1 q_1 }|^2\Big] \\
&\times (\delta f^q_{\p} f^q_{\k} (1-f^q_{\pp}) (1-f^q_{\kp})- (-\delta f^q_{\p}) (1-f^q_{\k}) f^q_{\pp} f^q_{\kp} \\
 &+f^q_{\p} \delta f^q_{\k} (1-f^q_{\pp}) (1-f^q_{\kp})- (1-f^q_{\p}) (-\delta f^q_{\k}) f^q_{\pp} f^q_{\kp} \\
 &+f^q_{\p} f^q_{\k} (-\delta f^q_{\pp}) (1-f^q_{\kp})- (1-f^q_{\p}) (1-f^q_{\k}) \delta f^q_{\pp} f^q_{\kp}\\
 &+f^q_{\p} f^q_{\k} (1-f^q_{\pp}) \delta q^g_{\kp}- (1-f^q_{\p}) (1-f^q_{\k}) f^q_{\pp} \delta f^q_{\kp})] \\
&\times \left[ \delta(\tilde{\p}-\p)\delta_{qs}+\delta(\tilde{\p}-\k)\delta_{qs}-\delta(\tilde{\p}-\pp)\delta_{qs}-\delta(\tilde{\p}-\kp)\delta_{qs}\right] \Bigg\}.
\end{aligned}
\end{equation}
We have explicitly summed over the different fermion flavours ($N_f = 3$) and over quarks and anti-quarks. 

The other contribution $\delta C^{s,(2)}_{2\leftrightarrow 2}[f](\ptilde)$ comes from the matrix elements that include the medium induced effective masses from \cref{eq:effective_mass_g,eq:effective_mass_q} in order to regulate soft momentum transfer, leading to
\begin{equation}
    \begin{aligned}
\delta C^{s,(2)}_{2\leftrightarrow 2}[f](\ptilde) &=  
\frac{1}{8\nu_{s}} (2\pi)^3\int_{\p\k\p'\k'}  \bigg\{ \Big[ \delta |\mathcal{M}^{gg}_{gg }|^2 \Big](f^g_{\p} f^g_{\k} (1+f^g_{\pp}) (1+f^g_{\kp})- (1+f^g_{\p}) 
 (1+f^g_{\k}) f^g_{\pp} f^g_{\kp})  \\
 &\times\left[ 
 \delta(\tilde{\p}-\p)\delta_{gs}+\delta(\tilde{\p}-\k)\delta_{gs}-\delta(\tilde{\p}-\pp)\delta_{g
  s}-\delta(\tilde{\p}-\kp)\delta_{g s}\right]\\
&+ \Big[2    \delta |\mathcal{M}^{gg}_{q_1\bar{q}_1}|^2\Big] (f^g_{\p} f^g_{\k} (1-f^q_{\pp}) 
(1-f^q_{\kp})- (1+f^g_{\p}) (1+f^g_{\k}) f^q_{\pp} f^q_{\kp}) \\
&\times \left[ (2 N_f) 
\delta(\tilde{\p}-\p)\delta_{gs}+(2 N_f) 
\delta(\tilde{\p}-\k)\delta_{gs}-\delta(\tilde{\p}-\pp)\delta_{qs}-\delta(\tilde{\p}-\kp)\delta_{q
 s}\right]\\
&+\Big[4  \delta |\mathcal{M}^{q_1 g}_{q_1 g}|^2\Big](f^q_{\p} f^g_{\k} (1-f^q_{\pp}) (1+f^g_{\kp})- 
(1-f^q_{\p}) (1+f^g_{\k}) f^q_{\pp} f^g_{\kp})\\
&\times \left[ \delta(\tilde{\p}-\p)\delta_{qs}+(2 
N_f)\delta(\tilde{\p}-\k)\delta_{gs}-\delta(\tilde{\p}-\pp)\delta_{qs}-(2 
N_f)\delta(\tilde{\p}-\kp)\delta_{g s}\right]\\
&+\Big[  2[2 (N_f-1)]\delta  |\mathcal{M}^{q_1 q_2}_{q_1 q_2}|^2+[2 (N_f-1)] 
\delta |\mathcal{M}^{q_1 \bar{q}_1}_{q_2\bar{q}_2}|^2+2  \delta |\mathcal{M}^{q_1 
\bar{q}_1}_{q_1\bar{q}_1}|^2+ \delta |\mathcal{M}^{q_1 q_1}_{q_1 q_1 }|^2\Big]\\
&\times(f^q_{\p} 
f^q_{\k} (1-f^q_{\pp}) (1-f^q_{\kp})- (1-f^q_{\p}) (1-f^q_{\k}) f^q_{\pp} f^q_{\kp}) ]\\
&\times\left[ 
\delta(\tilde{\p}-\p)\delta_{qs}+\delta(\tilde{\p}-\k)\delta_{qs}-\delta(\tilde{\p}-\pp)\delta_{qs}-\delta(\tilde{\p}-\kp)\delta_{q
 s}\right] \bigg\}.
    \end{aligned}
\end{equation}
The perturbed gluon mass is computed from
\begin{equation}\label{perturbed_mass}
    \delta m_g^2 = 2 g^2\int \frac{d^3\p}{(2\pi)^3 p}[N_c \delta f_g(p) + \frac{N_f}{2} (\delta f_q(p) + \delta f_{\Bar{q}}(p))],
\end{equation}
analogously $\delta m_q$. This results in the following perturbed matrix element, e.g., for $gq\leftrightarrow g\Bar{q}$
\begin{equation}
    \delta |\mathcal{M}^{gg}_{q_1 \bar{q}_1}|^2/g^4 = 8 d_F C_F \left[ C_F
    \frac{u-s}{t}\left(\frac{-\xi_g^2 \delta m_g^2}{q^2 + \xi_g^2 m_g^2}\right) 
    \right].
\end{equation}

\subsection{$1\leftrightarrow 2$-processes}

The inelastic kernel reads
\begin{equation}
\begin{aligned}
C^s_{1\leftrightarrow 2}[f](\ptilde)&=\frac{1}{2}
\frac{1}{\nu_s} \frac{(2\pi)^3}{4\pi\tilde{p}}
 \sum_{abc}\int_0^{\infty}dpdp'dk' 
 4\pi\gamma^{a}_{bc} 
 \delta(p-p'-k')\\
&\times \{ (f^a_{p\hat{\bm{n}}} (1\pm f^b_{p'\hat{\bm{n}}})(1\pm f^c_{k'\hat{\bm{n}}}))-(f^b_{p'\hat{\bm{n}}} 
f^c_{k'\hat{\bm{n}}} (1\pm f^a_{p\hat{\bm{n}}})\} \\
&\times\left[\delta(\Tilde{p}-p)\delta_{as}-\delta(\tilde{p}-p')\delta_{bs} 
-\delta(\tilde{p}-k')\delta_{cs}\right].
    \end{aligned}
\end{equation}
By factoring out the splitting functions we can write the rates $\gamma^{a}_{bc}$ for each channel as
\begin{align}
    \gamma^g_{gg}(p;p',k') &= \frac{p^4 + p'^4 + k'^4}{p^3p'^3k'^3}\mathcal{F}_g(p;p',k'),\\
        \gamma^q_{qg}(p;p',k') &= \frac{p^2 + p'^2}{p^2p'^2k'^3}\mathcal{F}_q(p;p',k'),\\
    \gamma^g_{qq}(p;p',k') &=\gamma^q_{qg}(k';-p',p),
\end{align}
where scatterings with the soft background are resummed resulting in an effective vertex~\cite{Arnold:2002zm}. $\p$ is the momentum of the parent parton that splits into two partons with momentum $\p' = x\p$ and $\k' = (1-x)\p$. The functions $\mathcal{F}_s(p;p',k')$ are given by
\begin{equation}
    \mathcal{F}_s(p;p',k') = \frac{\nu_sC_sg^2}{8(2\pi)^4}\int\frac{d^2\bm{h}}{(2\pi)^2}\bm{h}\cdot\mathrm{Re}\bm{F}_s(\bm{h};p,p',k'),
\end{equation}
where $\bm{F}_s$ is found by iteratively solving
\begin{equation}\label{eq:lpm_eq}
    \begin{aligned}
    2\bm{h} &= i\delta E(\bm{h})\bm{F}_s(\bm{h}) + g^2T_{*}\int \frac{d^2\bm{q_{\perp}}}{(2\pi)^2}\mathcal{A}(\bm{q}_{\perp})\\
    &\times \left \{ \frac{1}{2}(C_s + C_s - C_A)[\bm{F}_s(\bm{h}) - \bm{F}_s(\bm{h}-k'\bm{q}_{\perp})] \right.\\
    &\left.+ \frac{1}{2}(C_s + C_A - C_s)[\bm{F}_s(\bm{h}) - \bm{F}_s(\bm{h}-p'\bm{q}_{\perp})] \right.\\
    &\left.+ \frac{1}{2}(C_A + C_s - C_s)[\bm{F}_s(\bm{h}) - \bm{F}_s(\bm{h}-p\bm{q}_{\perp})] \right \}.
\end{aligned}
\end{equation}

The background fluctuations $\mathcal{A}(\bm{q}_{\perp})$ are treated in the isotropic screening approximation
\begin{equation}
    \mathcal{A}(\bm{q}_{\perp}) = \frac{1}{\bm{q}_{\perp}^2} - \frac{1}{\bm{q}_{\perp}^2 + 2m_g^2}.
\end{equation}
The energy difference $\delta E$ is given by
\begin{equation}
    \delta E(\bm{h}; p,p',k')\equiv \frac{m_g^2}{2k'} + \frac{m_s^2}{2p'} + \frac{m_s^2}{2p} + \frac{\bm{h}^2}{2pk'p'},
\end{equation}
and $T_*$ is the effective temperature
\begin{equation}
    T_*\equiv \frac{1}{\nu_g m_g^2}\sum_s\nu_s g^2 C_s \int\frac{d^3\p}{(2\pi)^3}f_s(\p)(1\pm f_s(\p)).
\end{equation}
We rewrite \cref{eq:lpm_eq} using the inverse Fourier transform 
\begin{equation}
    \bm{F}_s(\bm{b}) = \int\frac{d^2\bm{h}}{(2\pi)^2}e^{i\bm{b}\cdot\bm{h}}\bm{F}_s(\bm{h}),
\end{equation}
and the rescaled variables $\bm{b}\rightarrow\Tilde{\bm{b}}=m_gp\bm{b}$ and $\bm{F}\rightarrow\Tilde{\bm{F}}_s = \frac{1}{m_gp}\frac{1}{2pp'k'}\bm{F}_s$. From \cref{eq:lpm_eq}, we know that $\bm{F}_s \sim \bm{h}$. Since $\delta E$ is invariant under $\bm{h}\rightarrow - \bm{h}$, it follows that $\Tilde{\bm{F}}_s\sim \Tilde{\bm{b}}$. Therefore we can write $\Tilde{\bm{F}}_s(\Tilde{\bm{b}}) = \Tilde{\bm{b}}f_s(\Tilde{b})$ and obtain a Schrödinger-like equation for $f_s(\Tilde{b})$
\begin{equation}
\begin{aligned}
    0 &= \left( \partial_{\Tilde{b}^2} + \frac{3}{b}\partial_{\Tilde{b}} -\frac{M_s^2(x)}{m_g^2} \right)f_s(\Tilde{b})\\
    &+ i\eta  \frac{1}{2}\left( [2C_s -C_A]C((1-x)\Tilde{\bm{b}}) + C_A C(x\Tilde{\bm{b}}) + C_A C(\Tilde{\bm{b}}) \right)f_s(\Tilde{b}),
\end{aligned}
\end{equation}
where we have defined $\eta = x(1-x)\lambda T_*p/m_g$ and the effective mass
\begin{equation}
    M_s^2(x) = xm_g^2 + (1-x)^2m_s^2.
\end{equation}
The collision kernel $C$ is given by
\begin{align}\label{eq:schr_eq}
    C(\Tilde{\bm{b}}) &= \int\frac{d^2\bm{q}_{\perp}}{(2\pi)^2}\mathcal{A}(\bm{q}_\perp)\left[ 1 - e^{i\bm{q}_{\perp}\cdot \Tilde{\bm{b}}/m_g} \right]\\
    &= \frac{1}{2}\left[ K_0\left(\Tilde{b}\frac{m_D}{m_g}\right) + \gamma_E + \log\left(\Tilde{b}\frac{m_D}{2m_g}\right)\right].
\end{align}
The boundary conditions of \cref{eq:schr_eq} are given by~\cite{Kurkela:2022qhn}
\begin{align}
    f(\Tilde{b})&\overset{\Tilde{b}\rightarrow 0}{=}\frac{1}{\pi \Tilde{b}^2},\\
    f(\Tilde{b})&\overset{\Tilde{b}\rightarrow \infty}{=}0.
\end{align}
Solving for $f_s(\Tilde{b})$, we can relate it to $\mathcal{F}_s(p;p',k')$ via
\begin{equation}
    \mathcal{F}_s(p;p',k') = \frac{\nu_s C_sg^2}{(2\pi)^4}m_g^2p^3p'k'\mathrm{Im}f_s(0).
\end{equation}
In practice, \cref{eq:lpm_eq} is solved in two limits between which the solution is interpolated. In the Bethe-Heitler (BH) limit where the radiated gluon is very soft (small $x$ or $\eta \ll 1$), we obtain
\begin{equation}
\begin{aligned}
    \mathrm{Im}f_s(0) &= \frac{1}{C_A}\left[ (2C_s - C_A)\mathcal{Q}\left((1-x)\frac{m_D^2}{M_s^2(x)}\right) + C_A\mathcal{Q}\left(x\frac{m_D^2}{M_s^2(x)}\right) + C_A\mathcal{Q}\left( \frac{m_D^2}{M_s^2(x)}\right)\right],\\
    &\equiv f_{s,\mathrm{BH}}
    \end{aligned}
\end{equation}
with the function
\begin{equation}\label{eq:Qfunc}
    \mathcal{Q}(r) = \frac{1}{8\pi^2}\left[ \frac{i (r-2)\left[ \text{Li}_2\left(r_-\right) - 
\text{Li}_2\left(r_+\right)\right]}{ \sqrt{(4-r) r}}- \log (r)+2\right],
\end{equation}
where $r_{\pm}=1-\frac{r}{2}\pm\frac{i}{2} \sqrt{(4-r) r}$. Taking the opposite LPM-limit for $\eta \gg 1$ we get~\cite{Arnold:2008zu}
\begin{equation}
    \begin{aligned}
        \mathrm{Im}f_s(0) &= \frac{1}{8\pi^{3/2}}\frac{1}{C_A}\left\{\left( 2C_s-C_A\right)(1-x)^2 \ln\left(\frac{\xi \sqrt{2\eta}}{(1-x)^2} + 1\right)\right.\\
&\left.+C_A x^2 \ln\left(\frac{\xi \sqrt{2\eta}}{x^2} + 1\right)+C_A\ln\left(\frac{\xi \sqrt{2\eta}}{1} + 1\right)\right\}^{1/2},\\
&\equiv f_{s,\mathrm{LPM}}
    \end{aligned}
\end{equation}
with $\xi = e^{2-\gamma_E + \pi/4} \approx 9.09916$\footnote{Not to be confused with the anisotropy parameter in \cref{init_prl_distr}}. The interpolating function for $\mathrm{Im}f_s(0)$ can then be expressed as
\begin{equation}
    \mathrm{Im}f_s(0) = \left[\sqrt{\eta + 1} - 1\right]\left\{\frac{1}{1+\eta}f_{s,\mathrm{BH}} + \frac{\eta}{1+\eta}f_{s,\mathrm{LPM}}\right\}.
\end{equation}
Analogously to the elastic kernel $C_{2\leftrightarrow2}^s$, the inelastic kernel receives two contributions upon linearisation
\begin{equation}
    \delta C^{s}_{1\leftrightarrow 2}[f](\ptilde) = \delta C^{s,(1)}_{1\leftrightarrow 2}[f](\ptilde) + \delta C^{s,(2)}_{1\leftrightarrow 2}[f](\ptilde),
\end{equation}
where $\delta C^{s}_{1\leftrightarrow 2}[f](\ptilde)$ accounts for the perturbation of the loss and gain terms. For the linearisation of the rates $\gamma_{bc}^a$ in $\delta C^{s,(2)}_{1\leftrightarrow 2}[f](\ptilde)$ one takes into account
\begin{equation}
    \delta \eta = \eta\left( \frac{\delta T_*}{T_*} - \frac{\delta m_g^2}{m_g^2} \right).
\end{equation}
Additionally, in the BH-limit, one has to consider the perturbation of the effective masses $\delta M_s^2(x)$ which, in particular, only contribute to the channels that involve quarks.

\end{document}